\documentclass{article}
\usepackage{amssymb}
\usepackage{amsfonts}
\usepackage{amsmath}

\input{tcilatex}
\begin{document}

\title{ A spectral theorem for the semigroup generated by a class of
infinitely many master equations}
\author{Sabine B\"{o}gli$^{\ast }$ and Pierre-A. Vuillermot$^{\ast \ast
,\ast \ast \ast }$ \\
Department of Mathematical Sciences, Durham University, \\
Durham DH1 3LE, United Kingdom$^{\ast }$\\
UMR-CNRS 7502, Inst. \'{E}lie Cartan de Lorraine, Nancy, France$^{\ast \ast
} $\\
Grupo de F\'{\i}sica Matem\'{a}tica, GFMUL, Faculdade de Ci\^{e}ncias,\\
Universidade de Lisboa, 1749-016 Lisboa, Portugal$^{\ast \ast \ast }$}
\date{}
\maketitle

\begin{abstract}
In this article we investigate the spectral properties of the infinitesimal
generator of an infinite system of master equations arising in the analysis
of the approach to equilibrium in statistical mechanics. The system under
consideration thus consists of infinitely many first-order differential
equations governing the time evolution of probabilities susceptible of
describing jumps between the eigenstates of a differential operator with a
discrete point spectrum. The transition rates between eigenstates are chosen
in such a way that the so-called detailed balance conditions are satisfied,
so that for a large class of initial conditions the given system possesses a
global solution which converges exponentially rapidly toward a time
independent probability of Gibbs type. A particular feature and a challenge
of the problem under consideration is that in the infinite-dimensional
functional space where the initial-value problem is well posed, the
infinitesimal generator is realized as a non normal and non dissipative
compact operator, whose spectrum therefore does not exhibit a spectral gap
around the zero eigenvalue.
\end{abstract}

\section{Introduction and outline}

In the simplest setting a system of master equations refers to a set of
first-order linear ordinary differential equations which describe the time
evolution of probabilities susceptible of describing transitions between the
eigenstates of a given differential operator with a discrete point spectrum.
As such master equations play an important role in the analysis of certain
chemical reaction processes, radioactive decay processes and the propagation
of epidemics, to name only a few. More generally, they also allow one to
establish an important formal link between the laws that govern the
microscopic and reversible motion of particles of physical or chemical
systems and the macroscopic irreversible laws of thermodynamics, thereby
providing some understanding of the approach to equilibrium in statistical
mechanics (we refer the reader for instance to \cite{haake} and to Chapter V
in \cite{vankampen} for a history and many examples, and to \cite%
{mozgunovlidar}, \cite{tomeoliveira}, \cite{tomeoliveirabis} with their
plethora of references for recent advances in the subject of Stochastic
Thermodynamics). It is precisely this last aspect that we shall be concerned
with in this article, and accordingly we shall organize the remaining part
of this article in the following way: In Section 2 we start out with a
sequence of real numbers $\left( \lambda _{\mathsf{m}}\right) _{\mathsf{m}%
\in \mathbb{N}^{+}}$ chosen in such a way that the so-called partition
function satisfies%
\begin{equation}
Z_{\beta }:=\sum_{\mathsf{m=1}}^{\mathsf{+\infty }}\exp \left[ -\beta
\lambda _{\mathsf{m}}\right] <+\infty  \label{partitionfunction}
\end{equation}%
for all $\beta >0$, which implies in particular that $\lambda _{\mathsf{m}%
}\rightarrow +\infty $ as $\mathsf{m}\rightarrow +\infty $. We then define
the Gibbs probability vectors $\mathsf{p}_{\beta ,\mathsf{Gibbs}}$ by their
components%
\begin{equation}
p_{\beta ,\mathsf{m},\mathsf{Gibbs}}:=Z_{\beta }^{-1}\exp \left[ -\beta
\lambda _{\mathsf{m}}\right]  \label{gibbsbis}
\end{equation}%
for every $\mathsf{m}\in \mathbb{N}^{+}$, and with the sequence $\left(
\lambda _{\mathsf{m}}\right) _{\mathsf{m}\in \mathbb{N}^{+}}$ also consider
the class of initial-value problems for master equations of the form%
\begin{eqnarray}
\frac{dp_{\mathsf{m}}\left( \tau \right) }{d\tau } &=&\sum_{\mathsf{n=1}%
}^{+\infty }\left( r_{\mathsf{m,n}}p_{\mathsf{n}}\left( \tau \right) -r_{%
\mathsf{n,m}}p_{\mathsf{m}}\left( \tau \right) \right) ,\text{ \ }\tau \in %
\left[ 0,+\infty \right) ,  \notag \\
p_{\mathsf{m}}\left( 0\right) &=&p_{\mathsf{m}}^{\ast }
\label{masterequation}
\end{eqnarray}%
where $\left( p_{\mathsf{m}}^{\ast }\right) _{\mathsf{m\in }\mathbb{N}^{+}}$
stands for any sequence of initial-data satisfying%
\begin{equation}
p_{\mathsf{m}}^{\ast }\geq 0\text{, \ }\sum_{\mathsf{m=1}}^{+\infty }\text{\ 
}p_{\mathsf{m}}^{\ast }=1.  \label{probabilities}
\end{equation}%
In the preceding expression the time-independent transition rates $r_{%
\mathsf{m,n}}>0$ depend on $\lambda _{\mathsf{m}}$ and $\lambda _{\mathsf{n}%
} $ in a very specific way which we describe below, a choice that allows us
to prove the existence of a global solution to (\ref{masterequation}) and
provide a detailed investigation of the long-time behavior of each $p_{%
\mathsf{m}}\left( \tau \right) $ as $\tau \rightarrow +\infty $. More
specifically, by interpreting (\ref{masterequation}) as a dynamical system
defined in the usual Hilbert space $l_{\mathbb{C}}^{2}$ consisting of all
square summable complex sequences, we provide a complete spectral analysis
of the underlying infinitesimal generator $A$, which is realized there as a
non normal and non dissipative compact operator. In order to get compactness
we first show that $A$ is trace-class, then prove that $\nu _{1}=0$ is a
simple eigenvalue of $A$ whose eigenspace is generated by $\mathsf{p}_{\beta
,\mathsf{Gibbs}}$, proceed by proving that all the remaining eigenvalues are
simple, negative, and that the corresponding eigenvectors together with $%
\mathsf{p}_{\beta ,\mathsf{Gibbs}}$ constitute a complete system in $l_{%
\mathbb{C}}^{2}$ in the sense of Chapter V in \cite{gohbergkrein}. This is
done very indirectly as we have to show as a preliminary step that $A$ does
not possess any root vectors of height larger than one. Then, by imposing
additional constraints on the sequence $\left( \lambda _{\mathsf{m}}\right)
_{\mathsf{m}\in \mathbb{N}^{+}}$ and by proving some delicate estimates
related to the localization properties of the eigenvalues of $A$, we finally
prove that the complete system in question is actually a basis of $l_{%
\mathbb{C}}^{2}$. This eventually leads to the spectral decomposition of the
semigroup $\exp \left[ \tau A\right] _{\tau \in \left[ 0,+\infty \right) }$
generated by $A$, and thereby to the exponentially rapid convergence%
\begin{equation*}
p_{\mathsf{m}}\left( \tau \right) \rightarrow \gamma p_{\beta ,\mathsf{%
m,Gibbs}}
\end{equation*}%
for each $\mathsf{m}$ as $\tau \rightarrow +\infty $ for some $\gamma \in 
\mathbb{C}$ for a large class of initial data. The difficulty to be bypassed
in choosing such initial conditions is related to the fact that there is no
spectral gap between $\nu _{1}=0$ and the remaining eigenvalues since the
compactness of $A$ implies that $\nu _{1}=0$ is an accumulation point of the
spectrum, in contrast to finite-dimensional situations. We complete the
article with two appendices dealing respectively with some technical
question in the proof of Theorem 1 and with a geometric characterization of
the eigenvectors of the adjoint of $A$. Aside from its relation to
fundamental questions in Stochastic Thermodynamics, this paper was also
motivated by the desire to put our analysis into the perspective of the
spectral theory of non self-adjoint operators in Hilbert space as developed
in \cite{gohbergkrein}, which prompted us to choose $l_{\mathbb{C}}^{2}$ as
the underlying functional space where our analysis is carried out instead of
the geometrically less convenient Banach space $l_{\mathbb{C}}^{1}$
consisting of all absolutely summable complex sequences.

\section{A system of master equations as a dynamical system in $l_{\mathbb{C}%
}^{2}$}

We write $\left( .,.\right) _{2}$ for the usual inner product in $l_{\mathbb{%
C}}^{2}$, which we assume to be linear in the first argument and
complex-linear in the second, and $\left\Vert .\right\Vert _{2}$ for the
induced norm. Let us rewrite (\ref{masterequation}) as 
\begin{eqnarray}
\frac{dp_{\mathsf{m}}\left( \tau \right) }{d\tau } &=&\sum_{\mathsf{n=1}%
}^{+\infty }a_{\mathsf{m,n}}p_{\mathsf{n}}\left( \tau \right) ,\text{ \ }%
\tau \in \left[ 0,+\infty \right) ,  \notag \\
p_{\mathsf{m}}\left( 0\right) &=&p_{\mathsf{m}}^{\ast }\text{ \label%
{masterequationter}}
\end{eqnarray}%
with \ 
\begin{equation}
a_{\mathsf{m,n}}=\left\{ 
\begin{array}{c}
-\sum_{\mathsf{k=1},\text{ }\mathsf{k\neq m}}^{+\infty }r_{\mathsf{k,m}}%
\text{ \ \ for }\mathsf{m=n,} \\ 
\\ 
r_{\mathsf{m,n}}\text{\textsf{\ \ \ for }}\mathsf{m\neq n.}%
\end{array}%
\right.  \label{matrix}
\end{equation}%
Given a sequence $\left( \lambda _{\mathsf{m}}\right) _{\mathsf{m}\in 
\mathbb{N}^{+}}$ of real numbers satisfying (\ref{partitionfunction}), we
choose the transition rates from level $\mathsf{n}$ to level $\mathsf{m}$ in
(\ref{matrix}) as%
\begin{equation}
r_{\mathsf{m,n}}=c_{\mathsf{m,n}}\exp \left[ -\frac{\beta }{2}\left( \lambda
_{\mathsf{m}}-\lambda _{\mathsf{n}}\right) \right] ,  \label{transitionrate}
\end{equation}%
where the prefactors satisfy the symmetry condition $c_{\mathsf{m,n}}=c_{%
\mathsf{n,m}}$ for all $\mathsf{m,n}$\textsf{\ }$\in \mathbb{N}^{+}$ and are
otherwise arbitrary, so that the so-called detailed balance conditions%
\begin{equation}
r_{\mathsf{m,n}}p_{\beta ,\mathsf{n,Gibbs}}=r_{\mathsf{n,m}}p_{\beta ,%
\mathsf{m,Gibbs}}  \label{gibbs}
\end{equation}%
hold. We remark that the choice of (\ref{gibbsbis}) as initial data in (\ref%
{masterequationter}) then provides a time-independent solution to that
system of equations, and furthermore that conditions (\ref{gibbs}) also
ensure there is no entropy production in the system under consideration in
the sense of the definition proposed in \cite{schnakenberg} (see also
Section II in \cite{tomeoliveira} for more details). It is evidently clear,
however, that there are plenty of other choices than (\ref{transitionrate})
which satisfy (\ref{gibbs}) and for which the results of this article remain
valid, but we will keep using (\ref{transitionrate}) mainly for the sake of
clarity and simplification. Moreover, in order to make a well-defined
dynamical system in $l_{\mathbb{C}}^{2}$ out of (\ref{masterequationter}),
we shall use the freedom we have regarding the choice of the $c_{\mathsf{m,n}%
}$ by taking%
\begin{equation}
c_{\mathsf{m,n}}=\exp \left[ -\frac{\alpha \beta }{2}\left( \lambda _{%
\mathsf{m}}+\lambda _{\mathsf{n}}\right) \right]  \label{suitableconstants}
\end{equation}%
with $\alpha \in \left( 1,+\infty \right) $ for all $\mathsf{m,n}$\textsf{\ }%
$\in \mathbb{N}^{+}$, thereby obtaining the one-parameter family of
transition rates%
\begin{equation}
r_{\mathsf{m,n}}=\exp \left[ -\frac{1}{2}\left( \left( \alpha +1\right)
\lambda _{\mathsf{m}}+\left( \alpha -1\right) \lambda _{\mathsf{n}}\right)
\beta \right]  \label{rates}
\end{equation}%
indexed by $\alpha $. We observe that we then have%
\begin{equation}
\sum_{\mathsf{m=1}}^{+\infty }\sum_{\mathsf{n=1}}^{+\infty }r_{\mathsf{m,n}%
}^{2}=Z_{\left( \alpha -1\right) \beta }Z_{\left( \alpha +1\right) \beta
}<+\infty  \label{convergence1}
\end{equation}%
and%
\begin{equation}
\sum_{\mathsf{m=1}}^{+\infty }\left( \sum_{\mathsf{n=1}}^{+\infty }r_{%
\mathsf{n,m}}\right) ^{2}=Z_{\left( \alpha -1\right) \beta }Z_{\frac{\alpha
+1}{2}\beta }^{2}<+\infty  \label{convergence2}
\end{equation}%
according to (\ref{partitionfunction}).

\bigskip

The following preliminary result holds:

\bigskip

\textbf{Proposition 1. }\textit{Let us write }$\mathsf{p}=\left( p_{\mathsf{m%
}}\right) _{\mathsf{m\in }\mathbb{N}^{+}}$\textit{\ for any sequence in }$l_{%
\mathbb{C}}^{2}$\textit{. Then the expression}%
\begin{equation}
\left( A\mathsf{p}\right) _{\mathsf{m}}:=\sum_{\mathsf{n=1}}^{+\infty }a_{%
\mathsf{m,n}}p_{\mathsf{n}}  \label{operator}
\end{equation}%
\textit{defines a linear, non self-adjoint trace-class operator }$A:$\textit{%
\ }$l_{\mathbb{C}}^{2}\mapsto l_{\mathbb{C}}^{2}$\textit{\ whose trace is
given by}%
\begin{equation}
\func{Tr}A=Z_{\alpha \beta }-Z_{\frac{\alpha -1}{2}\beta }Z_{\frac{\alpha +1%
}{2}\beta }<0.  \label{trace}
\end{equation}

\bigskip

\textbf{Proof. }From (\ref{operator}) and the Cauchy-Schwarz inequality we
get%
\begin{equation*}
\left\Vert A\mathsf{p}\right\Vert _{2}^{2}\leq \sum_{\mathsf{m=1}}^{+\infty
}\sum_{\mathsf{n=1}}^{+\infty }\left\vert a_{\mathsf{m,n}}\right\vert
^{2}\times \left\Vert \mathsf{p}\right\Vert _{2}^{2}
\end{equation*}%
where%
\begin{eqnarray}
&&\sum_{\mathsf{m=1}}^{+\infty }\sum_{\mathsf{n=1}}^{+\infty }\left\vert a_{%
\mathsf{m,n}}\right\vert ^{2}  \notag \\
&=&\sum_{\mathsf{m=1}}^{+\infty }\left( \sum_{\mathsf{n=1},\text{ }\mathsf{%
n\neq m}}^{+\infty }r_{\mathsf{n,m}}\right) ^{2}+\sum_{\mathsf{m=1}%
}^{+\infty }\sum_{\mathsf{n=1},\text{ }\mathsf{n\neq m}}^{+\infty }r_{%
\mathsf{m,n}}^{2}  \label{convergenceagain} \\
&\leq &\sum_{\mathsf{m=1}}^{+\infty }\left( \sum_{\mathsf{n=1}}^{+\infty }r_{%
\mathsf{n,m}}\right) ^{2}+\sum_{\mathsf{m=1}}^{+\infty }\sum_{\mathsf{n=1}%
}^{+\infty }r_{\mathsf{m,n}}^{2}  \notag \\
&=&Z_{\left( \alpha -1\right) \beta }\left( Z_{\frac{\alpha +1}{2}\beta
}^{2}+Z_{\left( \alpha +1\right) \beta }\right) <+\infty  \notag
\end{eqnarray}%
according to (\ref{matrix}), (\ref{convergence1}) and (\ref{convergence2}),
so that $A$ is a linear bounded operator satisfying%
\begin{equation*}
\left\Vert A\mathsf{p}\right\Vert _{2}\leq c\left( \alpha ,\beta \right)
\left\Vert \mathsf{p}\right\Vert _{2}
\end{equation*}%
for every $\mathsf{p\in }l_{\mathbb{C}}^{2}$ for some $c\left( \alpha ,\beta
\right) >0$. The fact that $A$ is not self-adjoint in general is clear from
its definition. Therefore, in order to prove the trace-class property it is
necessary and sufficient to show that the series%
\begin{equation}
\sum_{\mathsf{m=1}}^{+\infty }\left( A\mathsf{h}_{\mathsf{m}},\mathsf{h}_{%
\mathsf{m}}\right) _{2}  \label{matrixtrace}
\end{equation}%
converges for any orthonormal basis $\left( \mathsf{h}_{\mathsf{m}}\right) _{%
\mathsf{m\in }\mathbb{N}^{+}}$ in $l_{\mathbb{C}}^{2}$, in which case the
value of (\ref{matrixtrace}) will not depend on the chosen basis (see, e.g.,
Section 2.3 in Chapter I of \cite{gelfandvilenkin} or more specifically
Theorem 8.1 in Chapter III of \cite{gohbergkrein}). In order to prove
convergence we introduce the canonical orthonormal basis in $l_{\mathbb{C}%
}^{2}$ defined by $\left( \mathsf{e}_{\mathsf{m}}\right) _{\mathsf{n}%
}=\delta _{\mathsf{m,n}}$ for all $\mathsf{m,n}\in \mathbb{N}^{+}$, and
expand each $\mathsf{h}_{\mathsf{m}}$ along that basis as%
\begin{equation*}
\mathsf{h}_{\mathsf{m}}=\sum_{\mathsf{n=1}}^{+\infty }\left( \mathsf{h}_{%
\mathsf{m}},\mathsf{e}_{\mathsf{n}}\right) _{2}\mathsf{e}_{\mathsf{n}}.
\end{equation*}%
Then we have%
\begin{equation*}
A\mathsf{h}_{\mathsf{m}}=\sum_{\mathsf{n=1}}^{+\infty }\left( \mathsf{h}_{%
\mathsf{m}},\mathsf{e}_{\mathsf{n}}\right) _{2}A\mathsf{e}_{\mathsf{n}}
\end{equation*}%
and therefore%
\begin{equation*}
\left( A\mathsf{h}_{\mathsf{m}},\mathsf{h}_{\mathsf{m}}\right) _{2}=\sum_{%
\mathsf{n=1}}^{+\infty }\sum_{\mathsf{k=1}}^{+\infty }a_{\mathsf{k,n}}\left( 
\mathsf{h}_{\mathsf{m}},\mathsf{e}_{\mathsf{n}}\right) _{2}\left( \mathsf{e}%
_{\mathsf{k}},\mathsf{h}_{\mathsf{m}}\right) _{2},
\end{equation*}%
so that we get%
\begin{eqnarray*}
&&\sum_{\mathsf{m=1}}^{+\infty }\left\vert \left( A\mathsf{h}_{\mathsf{m}},%
\mathsf{h}_{\mathsf{m}}\right) _{2}\right\vert \leq \frac{1}{2}\sum_{\mathsf{%
n=1}}^{+\infty }\sum_{\mathsf{k=1}}^{+\infty }\left\vert a_{\mathsf{k,n}%
}\right\vert \sum_{\mathsf{m=1}}^{+\infty }\left( \left\vert \left( \mathsf{h%
}_{\mathsf{m}},\mathsf{e}_{\mathsf{n}}\right) _{2}\right\vert
^{2}+\left\vert \left( \mathsf{h}_{\mathsf{m}},\mathsf{e}_{\mathsf{k}%
}\right) _{2}\right\vert ^{2}\right) \\
&=&\sum_{\mathsf{n=1}}^{+\infty }\sum_{\mathsf{k=1}}^{+\infty }\left\vert a_{%
\mathsf{k,n}}\right\vert \leq 2\sum_{\mathsf{n=1}}^{+\infty }\sum_{\mathsf{%
k=1}}^{+\infty }r_{\mathsf{k,n}}=2Z_{_{\frac{\alpha -1}{2}\beta }}Z_{\frac{%
\alpha +1}{2}\beta }<+\infty
\end{eqnarray*}%
since the relation%
\begin{equation*}
\sum_{\mathsf{m=1}}^{+\infty }\left\vert \left( \mathsf{h}_{\mathsf{m}},%
\mathsf{e}_{\mathsf{n}}\right) _{2}\right\vert ^{2}=\left\Vert \mathsf{e}_{%
\mathsf{n}}\right\Vert _{2}^{2}=1
\end{equation*}%
holds for every $\mathsf{n\in }\mathbb{N}^{+}$ as a consequence of the
expansion%
\begin{equation*}
\mathsf{e}_{\mathsf{n}}=\sum_{\mathsf{m=1}}^{+\infty }\left( \mathsf{e}_{%
\mathsf{n}},\mathsf{h}_{\mathsf{m}}\right) _{2}\mathsf{h}_{\mathsf{m}}.
\end{equation*}%
This proves the desired result with the actual value of the trace given by%
\begin{equation*}
\func{Tr}A=\sum_{\mathsf{m=1}}^{+\infty }\left( A\mathsf{e}_{\mathsf{m}},%
\mathsf{e}_{\mathsf{m}}\right) _{2}=-\sum_{\mathsf{m=1}}^{+\infty }\sum_{%
\mathsf{k=1},\text{ }\mathsf{k\neq m}}^{+\infty }r_{\mathsf{k,m}},
\end{equation*}%
which is (\ref{trace}) as a direct consequence of (\ref{partitionfunction})
and (\ref{rates}). \ \ $\blacksquare $

\bigskip

Aside from not being self-adjoint, $A$ is not a normal operator either as $%
AA^{\ast }\neq A^{\ast }A$ in general where $A^{\ast }$ stands for the
adjoint of $A$. Nor is it dissipative as the quadratic form of its imaginary
part fails to be positive. However, both $A$ and $A^{\ast }$ are compact as
trace class operators and we shall repeatedly use that property to prove the
results which follow. We begin with the following detailed description of
some spectral properties:

\bigskip

\textbf{Theorem 1. }\textit{Let }$A$\textit{\ be the operator defined by (%
\ref{operator}). Then the following statements hold:}

\textit{(a) The spectrum of }$A$,\textit{\ }$\sigma (A)$, \textit{is a
discrete compact set with infinitely many elements }$\left( \nu _{\mathsf{k}%
}\right) _{\mathsf{k}\in \mathbb{N}^{+}}$ \textit{which are all eigenvalues,
including }$\nu _{1}=0$.

\textit{(b) Assume in addition that }$\lambda _{\mathsf{m+1}}>\lambda _{%
\mathsf{m}}$ \textit{for every} $\mathsf{m}\in \mathbb{N}^{+}$. \textit{Then
each eigenvalue }$\nu _{\mathsf{k}}$ \textit{is implicitly characterized by} 
\textit{the relation}%
\begin{equation}
\dsum\limits_{\mathsf{m=1}}^{+\infty }\frac{\exp \left[ -\alpha \beta
\lambda _{\mathsf{m}}\right] }{\nu _{\mathsf{k}}+b_{\mathsf{m}}}=1
\label{characterization}
\end{equation}%
\textit{where}%
\begin{equation}
b_{\mathsf{m}}:=Z_{\frac{\alpha +1}{2}\beta }\exp \left[ -\frac{\alpha -1}{2}%
\beta \lambda _{\mathsf{m}}\right] .  \label{shorthand}
\end{equation}%
\textit{Moreover each such eigenvalue} \textit{is real, simple, and the
corresponding eigenspace is spanned by }$\mathsf{\hat{p}}_{\mathsf{k}%
}=\left( \hat{p}_{\mathsf{k,m}}\right) _{\mathsf{m}\in \mathbb{N}^{+}}$ 
\textit{where}%
\begin{equation}
\hat{p}_{\mathsf{k,m}}=\frac{\exp \left[ -\frac{\alpha +1}{2}\beta \lambda _{%
\mathsf{m}}\right] }{\nu _{\mathsf{k}}+b_{\mathsf{m}}}.  \label{generator}
\end{equation}%
\textit{In particular, the eigenspace associated with }$\nu _{1}=0$ \textit{%
is spanned by }$\mathsf{p}_{\beta ,\mathsf{Gibbs}}$\textit{. }

\textit{(c)} \textit{All the non-zero elements of }$\sigma (A)$\textit{\ are
negative. More specifically, under the same additional hypothesis as in
Statement (b) and if the eigenvalues are ordered in such a way that }$\nu _{%
\mathsf{k}}<$\textit{\ }$\nu _{\mathsf{k+1}}$ \textit{for every} $\mathsf{k}%
\in \left\{ 2,3,...\right\} $, \textit{then} \textit{we have the
localization property} $\nu _{\mathsf{k}}\in \left( -b_{\mathsf{k-1}},-b_{%
\mathsf{k}}\right) $ \textit{for every such} $\mathsf{k}$.

\bigskip

\textbf{Proof. }The very first part of Statement (a) follows from the fact
that $A$ is compact. We also have%
\begin{equation*}
A\mathsf{p}_{\beta ,\mathsf{Gibbs}}=0
\end{equation*}%
as a consequence of (\ref{gibbs}) where $\mathsf{p}_{\beta ,\mathsf{Gibbs}%
}\in l_{\mathbb{C}}^{2}$, so that $\nu _{1}=0$ is indeed an eigenvalue.

As for the proof of Statement (b), let us take $\mathsf{p}\in l_{\mathbb{C}%
}^{2}$ with $\mathsf{p\neq 0}$ and%
\begin{equation*}
A\mathsf{p}=\nu _{\mathsf{k}}\mathsf{p}.
\end{equation*}%
Owing to (\ref{matrix}) and (\ref{operator}), this is equivalent to
requiring that%
\begin{equation*}
\sum_{\mathsf{n=1}}^{\mathsf{+\infty }}r_{\mathsf{m,n}}p_{\mathsf{n}}-p_{%
\mathsf{m}}\sum_{\mathsf{n=1}}^{\mathsf{+\infty }}r_{\mathsf{n,m}}=\nu _{%
\mathsf{k}}p_{\mathsf{m}}
\end{equation*}%
for every $\mathsf{m}\in \mathbb{N}^{+}$. Therefore, using (\ref{rates}) we
obtain%
\begin{equation}
c_{\mathsf{p,}\alpha ,\beta }\exp \left[ -\frac{\alpha +1}{2}\beta \lambda _{%
\mathsf{m}}\right] =\left( \nu _{\mathsf{k}}+b_{\mathsf{m}}\right) p_{%
\mathsf{m}}  \label{eigenequation}
\end{equation}%
for every $\mathsf{m}$ after some rearrangements, where $b_{\mathsf{m}}$ is
given by (\ref{shorthand}) and 
\begin{equation}
c_{\mathsf{p,}\alpha ,\beta }:=\sum_{\mathsf{n=1}}^{\mathsf{+\infty }}\exp %
\left[ -\frac{\alpha -1}{2}\beta \lambda _{\mathsf{n}}\right] p_{\mathsf{n}}.
\label{shorthandbis}
\end{equation}%
Consequently, if $c_{\mathsf{p,}\alpha ,\beta }\neq 0$ then $\nu _{\mathsf{k}%
}+b_{\mathsf{m}}\neq 0$ for every $\mathsf{m}$ so that (\ref{eigenequation})
leads to%
\begin{equation}
\mathsf{p=}c_{\mathsf{p,}\alpha ,\beta }\mathsf{\hat{p}}_{\mathsf{k}},
\label{eigenvector}
\end{equation}%
and we claim that this is the only possible case. Indeed, on the one hand if 
$c_{\mathsf{p,}\alpha ,\beta }=0$ then $\left( \nu _{\mathsf{k}}+b_{\mathsf{m%
}}\right) p_{\mathsf{m}}=0$ for every $\mathsf{m}$. But on the other hand,
since $\mathsf{m\neq n}$ implies $\lambda _{\mathsf{m}}\neq $\textit{\ }$%
\lambda _{\mathsf{n}}$ we have $b_{\mathsf{m}}\neq $\textit{\ }$b_{\mathsf{n}%
}$, so that there may exist at most one $\mathsf{m}^{\ast }$ with $\nu _{%
\mathsf{k}}+b_{\mathsf{m}^{\ast }}=0$. If there is no such $\mathsf{m}^{\ast
}$ then we must have $p_{\mathsf{m}}=0$ for every $\mathsf{m}$, hence $%
\mathsf{p=0}$ which is not an eigenvector. If there is such an $\mathsf{m}%
^{\ast }$ then $p_{\mathsf{m}}=0$ for every $\mathsf{m\neq m}^{\ast }$ but
we may have $p_{\mathsf{m}^{\ast }}=0$ or $p_{\mathsf{m}^{\ast }}\neq 0$. In
the first case we get $\mathsf{p=0}$ once again, while in the second case (%
\ref{shorthandbis}) reduces to%
\begin{equation*}
c_{\mathsf{p,}\alpha ,\beta }=\exp \left[ -\frac{\alpha -1}{2}\beta \lambda
_{\mathsf{m}^{\ast }}\right] p_{\mathsf{m}^{\ast }}\neq 0,
\end{equation*}%
a contradiction. Therefore, the eigenspace associated with $\nu _{\mathsf{k}%
} $ is indeed the one-dimensional subspace generated by $\mathsf{\hat{p}}_{%
\mathsf{k}}$. The fact that $\mathsf{\hat{p}}_{\mathsf{k}}\in $ $l_{\mathbb{C%
}}^{2}$ is a simple consequence of (\ref{partitionfunction}) which is easily
verified in case of $\mathsf{p}_{\beta ,\mathsf{Gibbs}}$, while for $\nu _{%
\mathsf{k}}\neq 0$ we have%
\begin{equation*}
\dsum\limits_{\mathsf{m=1}}^{+\infty }\left\vert \nu _{\mathsf{k}}+b_{%
\mathsf{m}}\right\vert ^{2}\left\vert \hat{p}_{\mathsf{k,m}}\right\vert
^{2}=Z_{\left( \alpha +1\right) \beta }<+\infty
\end{equation*}%
from (\ref{partitionfunction}) and (\ref{generator}), which implies $\sum_{%
\mathsf{m=1}}^{+\infty }\left\vert \hat{p}_{\mathsf{k,m}}\right\vert
^{2}<+\infty $ by asymptotic comparison since $b_{\mathsf{m}}\rightarrow 0$
as $\mathsf{m\rightarrow +\infty }$. Finally, the substitution of (\ref%
{eigenvector}) into (\ref{shorthandbis}) using (\ref{generator}) gives (\ref%
{characterization}) whose imaginary part is then equal to zero, that is,%
\begin{equation*}
\dsum\limits_{\mathsf{m=1}}^{+\infty }\limfunc{Im}\frac{\exp \left[ -\alpha
\beta \lambda _{\mathsf{m}}\right] }{\nu _{\mathsf{k}}+b_{\mathsf{m}}}=-%
\limfunc{Im}\nu _{\mathsf{k}}\dsum\limits_{\mathsf{m=1}}^{+\infty }\frac{%
\exp \left[ -\alpha \beta \lambda _{\mathsf{m}}\right] }{\left\vert \nu _{%
\mathsf{k}}+b_{\mathsf{m}}\right\vert ^{2}}=0,
\end{equation*}%
which implies that each $\nu _{\mathsf{k}}$ is real.

As for Statement (c), let us first prove that $\nu _{\mathsf{k}}<0$ for
every non-zero\textit{\ }$\nu _{\mathsf{k}}\in \sigma (A)$. Since $\nu _{%
\mathsf{k}}$ is real the adjoint equation reads%
\begin{equation}
A^{\ast }\mathsf{q}=\nu _{\mathsf{k}}\mathsf{q\label{adjointoperator}}
\end{equation}%
for some $\mathsf{q}\in l_{\mathbb{C}}^{2}$ with $\mathsf{q\neq 0}$. Since $%
\mathsf{q}\in l_{\mathbb{C}}^{2}$ implies that $q_{\mathsf{m}}\rightarrow 0$
as $\mathsf{m\rightarrow +\infty }$ and since $\mathsf{q\neq 0}$, there
exists $\mathsf{m}^{\ast }\in \mathbb{N}^{+}$ such that $\left\vert q_{%
\mathsf{m}^{\ast }}\right\vert >0$ and $\left\vert q_{\mathsf{m}}\right\vert
\leqslant \left\vert q_{\mathsf{m}^{\ast }}\right\vert $ for every $\mathsf{m%
}$. Consequently, noting that (\ref{adjointoperator}) implies in particular
the relation%
\begin{equation*}
\left( \nu _{\mathsf{k}}-a_{\mathsf{m}^{\ast },\mathsf{m}^{\ast }}\right) q_{%
\mathsf{m}^{\ast }}=\sum_{\mathsf{n=1},\text{ }\mathsf{n\neq m}^{\ast
}}^{+\infty }r_{\mathsf{n,m}^{\ast }}q_{\mathsf{n}}
\end{equation*}%
according to (\ref{matrix}), we get after simplification 
\begin{equation}
\left\vert \left( \nu _{\mathsf{k}}-a_{\mathsf{m}^{\ast },\mathsf{m}^{\ast
}}\right) \right\vert \leqslant \sum_{\mathsf{n=1},\text{ }\mathsf{n\neq m}%
^{\ast }}^{+\infty }r_{\mathsf{n,m}^{\ast }}=\left\vert a_{\mathsf{m}^{\ast
},\mathsf{m}^{\ast }}\right\vert  \label{inequality}
\end{equation}%
or, equivalently,%
\begin{equation*}
\nu _{\mathsf{k}}^{2}-2a_{\mathsf{m}^{\ast },\mathsf{m}^{\ast }}\nu _{%
\mathsf{k}}\leqslant 0
\end{equation*}%
where $\nu _{\mathsf{k}}^{2}>0$ since $\nu _{\mathsf{k}}\neq 0$, and
therefore $\nu _{\mathsf{k}}<0$ because $a_{\mathsf{m}^{\ast },\mathsf{m}%
^{\ast }}<0$. More specifically, from our hypothesis regarding $\lambda _{%
\mathsf{m}}$ we get $b_{\mathsf{m}}>b_{\mathsf{m+1}}$ from (\ref{shorthand})
for every $\mathsf{m}\in \mathbb{N}^{+}$. Then we have $\nu _{\mathsf{k}}\in
\left( -b_{\mathsf{1}},0\right) $ for every $\mathsf{k}\in \left\{
2,3,...\right\} $, for if there were at least one $\mathsf{k}^{\ast }$ with $%
\nu _{\mathsf{k}^{\ast }}\notin \left( -b_{\mathsf{1}},0\right) $ we would
necessarily have $\nu _{\mathsf{k}^{\ast }}+b_{\mathsf{m}}\leqslant 0$ for
each $\mathsf{m}$, thereby contradicting (\ref{characterization}). Now, let
us consider the function $f:\left( -\infty ,0\right) \setminus \left\{ -b_{%
\mathsf{m}},\text{ }\mathsf{m}\in \mathbb{N}^{+}\right\} \mapsto \mathbb{R}$
given by%
\begin{equation}
f\left( \nu \right) :=\dsum\limits_{\mathsf{m=1}}^{\mathsf{+\infty }}\frac{%
\exp \left[ -\alpha \beta \lambda _{\mathsf{m}}\right] }{\nu +b_{\mathsf{m}}}%
.  \label{function}
\end{equation}%
Since $b_{\mathsf{m}}\rightarrow 0$ as $\mathsf{m\rightarrow +\infty }$ it
is plain that this series converges absolutely by asymptotic comparison and
by virtue of (\ref{partitionfunction}). Moreover, as a consequence of the
properties of the holomorphic continuation of (\ref{function}) investigated
in Appendix A, we have%
\begin{eqnarray*}
\lim_{\nu \searrow -b_{\mathsf{k-1}}}f\left( \nu \right) &=&+\infty , \\
\lim_{\nu \nearrow -b_{\mathsf{k}}}f\left( \nu \right) &=&-\infty
\end{eqnarray*}%
and $f^{\prime }\left( \nu \right) <0$ whenever $\nu \in \left( -b_{\mathsf{%
k-1}},-b_{\mathsf{k}}\right) $, so that in particular there exists a unique $%
\nu ^{\ast }\in \left( -b_{\mathsf{k-1}},-b_{\mathsf{k}}\right) $ with $%
f\left( \nu ^{\ast }\right) =1$. Therefore we necessarily have $\nu ^{\ast
}=\nu _{\mathsf{k}}$ as a consequence of (\ref{characterization}), which
proves the desired localization property. Finally we have $f\left( 0\right)
:=\lim_{\nu \nearrow 0}f\left( \nu \right) =1$, which is yet another way of
saying that (\ref{characterization}) also characterizes $\nu _{\mathsf{1}}=0$%
. \ \ $\blacksquare $

\bigskip

\textsc{Remarks.} (1) The eigenvectors of $A^{\ast }$ in (\ref%
{adjointoperator}) corresponding to $\nu _{\mathsf{k}}\neq 0$ can also be
determined by using the method that led to (\ref{generator}). Each one of
them is indeed a multiple of $\mathsf{\hat{q}}_{\mathsf{k}}$ whose
components are 
\begin{equation}
\hat{q}_{\mathsf{k,m}}=\frac{\exp \left[ -\frac{\alpha -1}{2}\beta \lambda _{%
\mathsf{m}}\right] }{\nu _{\mathsf{k}}+b_{\mathsf{m}}}  \label{generatorbis}
\end{equation}%
for each $\mathsf{k}\in \left\{ 2,3,...\right\} $ and every $\mathsf{m}\in 
\mathbb{N}^{+}$. Moreover, in contrast to $A$ it is interesting to note that 
$\nu _{\mathsf{1}}=0\in \sigma (A^{\ast })$ is not an eigenvalue.

(2) Whereas (\ref{characterization}) holds for all the eigenvalues of $A$,
the relation%
\begin{equation}
\dsum\limits_{\mathsf{m=1}}^{+\infty }\frac{\exp \left[ -\frac{\alpha +1}{2}%
\beta \lambda _{\mathsf{m}}\right] }{\nu _{\mathsf{k}}+b_{\mathsf{m}}}=0
\label{characterizationter}
\end{equation}%
only characterizes the non-zero eigenvalues, that is, holds for each $%
\mathsf{k}\in \left\{ 2,3,...\right\} $. Indeed for every such $\mathsf{k}$
we have 
\begin{eqnarray*}
0 &=&f\left( \nu _{\mathsf{k}}\right) -f\left( 0\right) \\
&=&-\nu _{\mathsf{k}}\dsum\limits_{\mathsf{m=1}}^{+\infty }\frac{\exp \left[
-\alpha \beta \lambda _{\mathsf{m}}\right] }{b_{\mathsf{m}}\left( \nu _{%
\mathsf{k}}+b_{\mathsf{m}}\right) } \\
&=&-\frac{\nu _{\mathsf{k}}}{Z_{\frac{\alpha +1}{2}\beta }}\dsum\limits_{%
\mathsf{m=1}}^{+\infty }\frac{\exp \left[ -\frac{\alpha +1}{2}\beta \lambda
_{\mathsf{m}}\right] }{\nu _{\mathsf{k}}+b_{\mathsf{m}}}
\end{eqnarray*}%
from (\ref{characterization}), (\ref{shorthand}) and (\ref{function}), which
leads to (\ref{characterizationter}) since $\nu _{\mathsf{k}}\neq 0$.
Relation (\ref{characterizationter}) will play an important role later on
when we prove that the $\mathsf{\hat{p}}_{\mathsf{k}}$ constitute a basis of 
$l_{\mathbb{C}}^{2}$ under a more stringent condition regarding the sequence 
$\left( \lambda _{\mathsf{m}}\right) _{\mathsf{m}\in \mathbb{N}^{+}}$, a
fact that will lead to the spectral decomposition of the semigroup generated
by $A$ and thereby to its ultimate behavior for large times.

\bigskip

For now our objective is to prove the completeness of the $\mathsf{\hat{p}}_{%
\mathsf{k}}$ in the sense that the set of all their finite linear
combinations is everywhere dense in $l_{\mathbb{C}}^{2}$, provided we impose
an additional restriction on the parameter $\alpha $. Our preliminary step
in that direction is to rule out the existence of root vectors of height
larger than one. Let us recall that a root vector $\mathsf{q}\in $ $l_{%
\mathbb{C}}^{2}$, $\mathsf{q\neq 0}$, associated with the eigenvalue $\nu _{%
\mathsf{k}}$ of $A$ is one that satisfies%
\begin{equation*}
\left( A-\nu _{\mathsf{k}}\right) ^{\mathsf{n}}\mathsf{q=0}
\end{equation*}%
for some $\mathsf{n}\in \mathbb{N}^{+}$, and that the height $\mathfrak{h}(%
\mathsf{q})$ of $\mathsf{q}$ is the least integer for which the preceding
relation holds (see, e.g., Chapter 6 in \cite{vinberg} for a general
definition). The precise result is the following:

\bigskip

\textbf{Proposition 2.} \textit{Let us} \textit{assume that }$\lambda _{%
\mathsf{m+1}}>\lambda _{\mathsf{m}}$ \textit{for every} $\mathsf{m}\in 
\mathbb{N}^{+}$. \textit{Then we have}%
\begin{equation}
\ker \left( A-\nu _{\mathsf{k}}\right) ^{\mathsf{n}}=\ker \left( A-\nu _{%
\mathsf{k}}\right) =\func{span}\left\{ \mathsf{\hat{p}}_{\mathsf{k}}\right\}
\label{equalkernels}
\end{equation}%
\textit{for every} $\mathsf{n}\in \mathbb{N}^{+}$ \textit{and every} $%
\mathsf{k}\in \mathbb{N}^{+}$, \textit{where} $\mathsf{\hat{p}}_{\mathsf{k}}$
\textit{is given by (\ref{generator}). Thus }$\mathfrak{h}(\mathsf{q})=1$ 
\textit{for every root vector of }$A$.

\bigskip

\textbf{Proof.} The statement is true for $\mathsf{n=1}$ according to (b) of
Theorem 1, so that we proceed by induction on $\mathsf{n}$. Assuming then
that (\ref{equalkernels}) holds we take $\mathsf{q}\in \ker \left( A-\nu _{%
\mathsf{k}}\right) ^{\mathsf{n+1}}$, which gives%
\begin{equation*}
\left( A-\nu _{\mathsf{k}}\right) ^{\mathsf{n+1}}\mathsf{q=}\left( A-\nu _{%
\mathsf{k}}\right) ^{\mathsf{n}}\left( A-\nu _{\mathsf{k}}\right) \mathsf{q=0%
}
\end{equation*}%
and hence 
\begin{equation}
\left( A-\nu _{\mathsf{k}}\right) \mathsf{q=\gamma \hat{p}}_{\mathsf{k}}
\label{rangecondition}
\end{equation}%
for some $\gamma \in \mathbb{C}$. We proceed to show that $\gamma =0$ is the
only value for which (\ref{rangecondition}) may hold by proving that there
is no $\mathsf{q}\in $ $l_{\mathbb{C}}^{2}$ with%
\begin{equation}
\left( A-\nu _{\mathsf{k}}\right) \mathsf{q=\hat{p}}_{\mathsf{k}}.
\label{rangeconditionbis}
\end{equation}%
Arguing indirectly and using the same method as in the proof of Theorem 1 we
see that (\ref{rangeconditionbis}) is equivalent to having%
\begin{equation}
q_{\mathsf{k},\mathsf{m}}=c_{\mathsf{q,}\alpha ,\beta }\hat{p}_{\mathsf{k,m}%
}-\frac{\hat{p}_{\mathsf{k,m}}}{\nu _{\mathsf{k}}+b_{\mathsf{m}}}
\label{components}
\end{equation}%
for every $\mathsf{m}\in \mathbb{N}^{+}$, where we used (\ref{generator}) and%
\begin{equation}
c_{\mathsf{q,}\alpha ,\beta }:=\sum_{\mathsf{n=1}}^{\mathsf{+\infty }}\exp %
\left[ -\frac{\alpha -1}{2}\beta \lambda _{\mathsf{n}}\right] q_{\mathsf{n}}%
\text{.}  \label{shorthandter}
\end{equation}%
If such a $\mathsf{q}$ were to provide a solution to (\ref{rangeconditionbis}%
), its components (\ref{components}) should be compatible with $c_{\mathsf{q,%
}\alpha ,\beta }$ given by (\ref{shorthandter}). But the substitution of (%
\ref{components}) into (\ref{shorthandter}) and a repeated use of (\ref%
{generator}) along with (\ref{characterization}) show that compatibility is
possible if, and only if,%
\begin{equation*}
\sum_{\mathsf{m=1}}^{\mathsf{+\infty }}\frac{\exp \left[ -\alpha \beta
\lambda _{\mathsf{m}}\right] }{\left( \nu _{\mathsf{k}}+b_{\mathsf{m}%
}\right) ^{2}}=0
\end{equation*}%
for every $\mathsf{k}\in \mathbb{N}^{+}$, which contradicts the fact that
each term of the preceding expression is positive. Therefore (\ref%
{rangecondition}) is only valid with $\gamma =0$ so that%
\begin{equation*}
\ker \left( A-\nu _{\mathsf{k}}\right) ^{\mathsf{n+1}}\subseteq \ker \left(
A-\nu _{\mathsf{k}}\right) ,
\end{equation*}%
which proves the desired result since the converse inclusion is trivial. \ \ 
$\blacksquare $

\bigskip

\textsc{Remark.} With an identical proof we get a similar result for $%
A^{\ast }$, namely,%
\begin{equation}
\ker \left( A^{\ast }-\nu _{\mathsf{k}}\right) ^{\mathsf{n}}=\ker \left(
A^{\ast }-\nu _{\mathsf{k}}\right) =\func{span}\left\{ \mathsf{\hat{q}}_{%
\mathsf{k}}\right\}  \label{equalkernelsbis}
\end{equation}%
for every $\mathsf{n}\in \mathbb{N}^{+}$ and every $\mathsf{k}\in \left\{
2,3,...\right\} $, where $\mathsf{\hat{q}}_{\mathsf{k}}$ is given by (\ref%
{generatorbis}).

\bigskip

There are many known and well-documented criteria that ensure the
completeness of the root vectors of a given non self-adjoint operator on a
Hilbert space, and thereby the possibility of constructing a basis
consisting of such vectors (see, e.g., Chapters V and VI in \cite%
{gohbergkrein}). As far as $A$ is concerned we shall settle for an
application of a theorem that originally appeared in \cite{keldys}, which is
stated and proved as Theorem 8.1 in Chapter V of \cite{gohbergkrein}. Our
application, however, will be very indirect given the fact that the theorem
in question requires the operator under investigation to have a trivial
kernel, a property not shared by the operator $A$. Nevertheless, we will now
show that we can bypass this difficulty by means of an auxiliary operator
that has the desired properties. Let us denote by%
\begin{equation*}
P:l_{\mathbb{C}}^{2}\mapsto \ker A
\end{equation*}%
the orthogonal projection onto the eigenspace generated by $\mathsf{p}%
_{\beta ,\mathsf{Gibbs}}$, and let us consider the compact operator $A+P$
whose eigenvalues we denote by $\left( \varkappa _{\mathsf{k}}\right) _{%
\mathsf{k}=1}^{+\infty }$. In the sequel and for the sake of convenience we
write $E_{\nu _{\mathsf{k}}}(A)$ for the eigenspace of the operator $A$
associated with the eigenvalue $\nu _{\mathsf{k}}$, $E_{\varkappa _{\mathsf{k%
}}}(A+P)$ for that of the operator $A+P$ associated with the eigenvalue $%
\varkappa _{\mathsf{k}}$ and%
\begin{equation*}
R_{\varkappa _{\mathsf{k}}}(A+P)=\dbigcup\limits_{\mathsf{n=1}}^{+\infty
}\ker \left( A+P-\varkappa _{\mathsf{k}}\right) ^{\mathsf{n}}
\end{equation*}%
for the corresponding root vector subspace.

We have the following preliminary result:

\bigskip

\textbf{Proposition 3.} \textit{Let us assume that }$\lambda _{\mathsf{m+1}%
}>\lambda _{\mathsf{m}}$ \textit{for every} $\mathsf{m}\in \mathbb{N}^{+}$. 
\textit{Then we have}%
\begin{equation}
\ker \left( A+P\right) =\left\{ 0\right\}  \label{kernelbis}
\end{equation}%
\textit{and }%
\begin{equation}
\dbigvee\limits_{\mathsf{k=1}}^{+\infty }E_{\nu _{\mathsf{k}%
}}(A)=\dbigvee\limits_{\mathsf{k=1}}^{+\infty }R_{\varkappa _{\mathsf{k}%
}}(A+P),  \label{spaces}
\end{equation}%
\textit{where the spaces in (\ref{spaces}) stand for the closed linear hull
of }$\cup _{\mathsf{k=1}}^{+\infty }E_{\nu _{\mathsf{k}}}(A)$ \textit{and} $%
\cup _{\mathsf{k=1}}^{+\infty }R_{\varkappa _{\mathsf{k}}}(A+P)$, \textit{%
respectively.}

\bigskip

\textbf{Proof. }If $\mathsf{q}\in \ker \left( A+P\right) $ we have%
\begin{equation}
A\mathsf{q=-}\left( \mathsf{q,\hat{p}}_{\beta ,\mathsf{Gibbs}}\right) _{2}%
\mathsf{\hat{p}}_{\beta ,\mathsf{Gibbs}}  \label{relation}
\end{equation}%
where $\mathsf{\hat{p}}_{\beta ,\mathsf{Gibbs}}$ now stands for $\mathsf{p}%
_{\beta ,\mathsf{Gibbs}}$ renormalized in such a way that $\left\Vert 
\mathsf{\hat{p}}_{\beta ,\mathsf{Gibbs}}\right\Vert _{2}=1$, and therefore $%
\mathsf{q}\in E_{\nu _{\mathsf{1}}=0}(A)^{\perp }$ as a consequence of the
non existence result of a $\mathsf{q}$ satisfying (\ref{rangecondition})
when $\gamma \neq 0$. On the other hand we infer from (\ref{relation}) that $%
A^{2}\mathsf{q=0}$, which implies $\mathsf{q}\in E_{\nu _{\mathsf{1}}=0}(A)$
by virtue of Proposition 2 and thereby $\mathsf{q}\in E_{\nu _{\mathsf{1}%
}=0}(A)\cap E_{\nu _{\mathsf{1}}=0}(A)^{\perp }$ =$\left\{ 0\right\} $,
which proves (\ref{kernelbis}).

Next, we show that 
\begin{equation}
\dbigvee\limits_{\mathsf{k=1}}^{+\infty }E_{\nu _{\mathsf{k}}}(A)\subseteq
\dbigvee\limits_{\mathsf{k=1}}^{+\infty }R_{\varkappa _{\mathsf{k}}}(A+P).
\label{inclusionbis}
\end{equation}%
We first have 
\begin{equation}
E_{\nu _{\mathsf{1}}=0}(A)\subseteq \dbigvee\limits_{\mathsf{k=1}}^{+\infty
}R_{\varkappa _{\mathsf{k}}}(A+P)  \label{inclusion}
\end{equation}%
since $\mathsf{\hat{p}}_{\beta ,\mathsf{Gibbs}}$ is also an eigenvector of $%
A+P$ with eigenvalue $\varkappa =1$, that is,%
\begin{equation}
\left( A+P\right) \mathsf{\hat{p}}_{\beta ,\mathsf{Gibbs}}=\mathsf{\hat{p}}%
_{\beta ,\mathsf{Gibbs}}.  \label{eigenequationbis}
\end{equation}%
Moreover, all the non zero eigenvalues of $A$ are eigenvalues of $A+P$ for
some suitably constructed eigenvector. Indeed, for each $\mathsf{k}\in
\left\{ 2,3...\right\} $ let us pick an arbitrary $\mathsf{q}_{\mathsf{k}%
}\in E_{\nu _{\mathsf{k}}}(A)$ with $\mathsf{q}_{\mathsf{k}}\neq 0$, and let 
$\mathsf{q}_{\mathsf{k},\gamma }=\mathsf{q}_{\mathsf{k}}+\gamma \mathsf{\hat{%
p}}_{\beta ,\mathsf{Gibbs}}$ for $\gamma \in \mathbb{C}$. Using the
properties of $A$ and $P$ already established we then get%
\begin{equation*}
\left( A+P-\nu _{\mathsf{k}}\right) \mathsf{q}_{\mathsf{k},\gamma }=P\mathsf{%
q}_{\mathsf{k}}+\gamma \left( 1-\nu _{\mathsf{k}}\right) \mathsf{\hat{p}}%
_{\beta ,\mathsf{Gibbs}}
\end{equation*}%
where%
\begin{equation*}
P\mathsf{q}_{\mathsf{k}}=\left( \mathsf{q}_{\mathsf{k}}\mathsf{,\hat{p}}%
_{\beta ,\mathsf{Gibbs}}\right) _{2}\mathsf{\hat{p}}_{\beta ,\mathsf{Gibbs}}%
\text{.}
\end{equation*}%
Thus%
\begin{equation*}
\left( A+P-\nu _{\mathsf{k}}\right) \mathsf{q}_{\mathsf{k},\gamma }=0
\end{equation*}%
if, and only if,%
\begin{equation*}
\left( \left( \mathsf{q}_{\mathsf{k}}\mathsf{,\hat{p}}_{\beta ,\mathsf{Gibbs}%
}\right) _{2}+\gamma \left( 1-\nu _{\mathsf{k}}\right) \right) \mathsf{\hat{p%
}}_{\beta ,\mathsf{Gibbs}}=0.
\end{equation*}%
But according to Statement (b) of Theorem 1 we have $\nu _{\mathsf{k}}\neq 1$
for every $\mathsf{k}$, so that the choice of 
\begin{equation*}
\gamma _{\mathsf{k}}=\frac{\left( \mathsf{q}_{\mathsf{k}}\mathsf{,\hat{p}}%
_{\beta ,\mathsf{Gibbs}}\right) _{2}}{\nu _{\mathsf{k}}-1}
\end{equation*}%
leads to $\mathsf{\hat{q}}_{\mathsf{k}}:=\mathsf{q}_{\mathsf{k}}+\gamma _{%
\mathsf{k}}\mathsf{\hat{p}}_{\beta ,\mathsf{Gibbs}}\in E_{\nu _{\mathsf{k}%
}}(A+P)$ with $\mathsf{\hat{q}}_{\mathsf{k}}\neq 0$. Indeed, $\mathsf{\hat{q}%
}_{\mathsf{k}}=0$ would imply $\mathsf{q}_{\mathsf{k}}\in E_{\nu _{\mathsf{1}%
}=0}(A),$ a contradiction. Consequently, because of (\ref{eigenequationbis})
we have \textsf{both} $\mathsf{\hat{q}}_{\mathsf{k}},\mathsf{\hat{p}}_{\beta
,\mathsf{Gibbs}}\in \cup _{\mathsf{k=1}}^{+\infty }R_{\varkappa _{\mathsf{k}%
}}(A+P)$, of which $\mathsf{q}_{\mathsf{k}}=\mathsf{\hat{q}}_{\mathsf{k}%
}-\gamma _{\mathsf{k}}\mathsf{\hat{p}}_{\beta ,\mathsf{Gibbs}}$ is a linear
combination. This along with (\ref{inclusion}) proves that%
\begin{equation*}
E_{\nu _{\mathsf{k}}}(A)\subseteq \dbigvee\limits_{\mathsf{k=1}}^{+\infty
}R_{\varkappa _{\mathsf{k}}}(A+P)
\end{equation*}%
for every $\mathsf{k}\in \mathbb{N}^{+}$, from which (\ref{inclusionbis})
follows.

In order to prove the converse inclusion we first observe that%
\begin{equation}
E_{\varkappa =1}\left( A+P\right) \subseteq \dbigvee\limits_{\mathsf{k=1}%
}^{+\infty }E_{\nu _{\mathsf{k}}}(A).  \label{inclusionter}
\end{equation}%
It follows indeed from the projection theorem in Hilbert space that%
\begin{equation*}
E_{\varkappa =1}\left( A+P\right) =E_{\nu _{1}=0}(A),
\end{equation*}%
for the relation $\left( A+P-1\right) \mathsf{q=0}$ with $\mathsf{q=\gamma 
\hat{p}}_{\beta ,\mathsf{Gibbs}}+\mathsf{\hat{q}}$ for some $\gamma \in 
\mathbb{C}$ and $\left( \mathsf{\hat{q},\hat{p}}_{\beta ,\mathsf{Gibbs}%
}\right) _{2}=0$ implies that $A\mathsf{\hat{q}=\hat{q}}$, hence that $%
\mathsf{\hat{q}}=0$ since $\nu =1$ is not an eigenvalue of $A$, which leads
to (\ref{inclusionter}).

For the other eigenvalues of $A+P$ let us take an arbitray $\mathsf{q}_{%
\mathsf{k}}\in R_{\varkappa _{\mathsf{k}}}(A+P)$. Then there exists $\mathsf{%
n}^{\ast }\in \mathbb{N}^{+}$ such that 
\begin{equation}
\left( A+P-\varkappa _{\mathsf{k}}\right) ^{\mathsf{n}^{\ast }}\mathsf{q}_{%
\mathsf{k}}=0.  \label{kernelter}
\end{equation}%
Moreover, since $AP=0$ we have the operator equality%
\begin{equation}
\left( A+P-\varkappa _{\mathsf{k}}\right) ^{\mathsf{n}}=\left( A-\varkappa _{%
\mathsf{k}}\right) ^{\mathsf{n}}+P\dsum\limits_{\mathsf{j=0}}^{\mathsf{n-1}%
}\left( 1-\varkappa _{\mathsf{k}}\right) ^{\mathsf{n-1-j}}\left( A-\varkappa
_{\mathsf{k}}\right) ^{\mathsf{j}}  \label{operatorequality}
\end{equation}%
valid for every $\mathsf{n}\in \mathbb{N}^{+}$, which follows from an easy
induction argument. We then proceed as in the first part of the proof by
considering $\mathsf{q}_{\mathsf{k},\gamma }=\mathsf{q}_{\mathsf{k}}+\gamma 
\mathsf{\hat{p}}_{\beta ,\mathsf{Gibbs}}$ for $\gamma \in \mathbb{C}$, and
by determining the result of the action of both sides of (\ref%
{operatorequality}) on $\mathsf{q}_{\mathsf{k},\gamma }$ when $\mathsf{n}=%
\mathsf{n}^{\ast }$. For the left-hand side we have%
\begin{equation}
\left( A+P-\varkappa _{\mathsf{k}}\right) ^{\mathsf{n}^{\ast }}\mathsf{q}_{%
\mathsf{k},\gamma }=\gamma \left( 1-\varkappa _{\mathsf{k}}\right) ^{\mathsf{%
n}^{\ast }}\mathsf{\hat{p}}_{\beta ,\mathsf{Gibbs}}  \label{relationbis}
\end{equation}%
as a consequence of (\ref{kernelter}), while for the right-hand side we may
write%
\begin{eqnarray}
&&\left( A-\varkappa _{\mathsf{k}}\right) ^{\mathsf{n}^{\ast }}\mathsf{q}_{%
\mathsf{k},\gamma }+P\dsum\limits_{\mathsf{j=0}}^{\mathsf{n}^{\ast }\mathsf{%
-1}}\left( 1-\varkappa _{\mathsf{k}}\right) ^{\mathsf{n}^{\ast }\mathsf{-1-j}%
}\left( A-\varkappa _{\mathsf{k}}\right) ^{\mathsf{j}}\mathsf{q}_{\mathsf{k}%
,\gamma }  \notag \\
&=&\left( A-\varkappa _{\mathsf{k}}\right) ^{\mathsf{n}^{\ast }}\mathsf{q}_{%
\mathsf{k},\gamma }+P\dsum\limits_{\mathsf{j=0}}^{\mathsf{n}^{\ast }\mathsf{%
-1}}\left( 1-\varkappa _{\mathsf{k}}\right) ^{\mathsf{n}^{\ast }\mathsf{-1-j}%
}\left( A-\varkappa _{\mathsf{k}}\right) ^{\mathsf{j}}\mathsf{q}_{\mathsf{k}}
\label{relationter} \\
&&+\gamma \dsum\limits_{\mathsf{j=0}}^{\mathsf{n}^{\ast }\mathsf{-1}}\left(
1-\varkappa _{\mathsf{k}}\right) ^{\mathsf{n}^{\ast }\mathsf{-1-j}}\left(
-\varkappa _{\mathsf{k}}\right) ^{\mathsf{j}}\mathsf{\hat{p}}_{\beta ,%
\mathsf{Gibbs}}  \notag
\end{eqnarray}%
since%
\begin{equation*}
P\left( A-\varkappa _{\mathsf{k}}\right) ^{\mathsf{j}}\mathsf{\hat{p}}%
_{\beta ,\mathsf{Gibbs}}=\left( -\varkappa _{\mathsf{k}}\right) ^{\mathsf{j}}%
\mathsf{\hat{p}}_{\beta ,\mathsf{Gibbs}}
\end{equation*}%
for every $\mathsf{j}$. But (\ref{relationbis}) and (\ref{relationter}) are
equal, so that by regrouping and rearranging terms we obtain%
\begin{equation*}
\left( A-\varkappa _{\mathsf{k}}\right) ^{\mathsf{n}^{\ast }}\mathsf{q}_{%
\mathsf{k},\gamma }=\gamma \left( -\varkappa _{\mathsf{k}}\right) ^{\mathsf{n%
}^{\ast }}\mathsf{\hat{p}}_{\beta ,\mathsf{Gibbs}}-P\dsum\limits_{\mathsf{j=0%
}}^{\mathsf{n}^{\ast }\mathsf{-1}}\left( 1-\varkappa _{\mathsf{k}}\right) ^{%
\mathsf{n}^{\ast }\mathsf{-1-j}}\left( A-\varkappa _{\mathsf{k}}\right) ^{%
\mathsf{j}}\mathsf{q}_{\mathsf{k}}
\end{equation*}%
where we have used the identity%
\begin{equation*}
\dsum\limits_{\mathsf{j=0}}^{\mathsf{n-1}}\left( 1-\varkappa _{\mathsf{k}%
}\right) ^{\mathsf{n-1-j}}\left( -\varkappa _{\mathsf{k}}\right) ^{\mathsf{j}%
}-\left( 1-\varkappa _{\mathsf{k}}\right) ^{\mathsf{n}}=-\left( -\varkappa _{%
\mathsf{k}}\right) ^{\mathsf{n}}
\end{equation*}%
valid for every $\mathsf{n}\in \mathbb{N}^{+}$. Thus we have%
\begin{equation*}
\left( A-\varkappa _{\mathsf{k}}\right) ^{\mathsf{n}^{\ast }}\mathsf{q}_{%
\mathsf{k},\gamma }=0
\end{equation*}%
if, and only if, 
\begin{equation*}
\gamma \left( -\varkappa _{\mathsf{k}}\right) ^{\mathsf{n}^{\ast }}\mathsf{%
\hat{p}}_{\beta ,\mathsf{Gibbs}}=P\dsum\limits_{\mathsf{j=0}}^{\mathsf{n}%
^{\ast }\mathsf{-1}}\left( 1-\varkappa _{\mathsf{k}}\right) ^{\mathsf{n}%
^{\ast }\mathsf{-1-j}}\left( A-\varkappa _{\mathsf{k}}\right) ^{\mathsf{j}}%
\mathsf{q}_{\mathsf{k}}.
\end{equation*}%
But according to (\ref{kernelbis}) we have $\varkappa _{\mathsf{k}}\neq 0$
for every $\mathsf{k}$, so that we may choose%
\begin{equation*}
\gamma _{\mathsf{k}}=\left( -\varkappa _{\mathsf{k}}\right) ^{-\mathsf{n}%
^{\ast }}\left( P\dsum\limits_{\mathsf{j=0}}^{\mathsf{n}^{\ast }\mathsf{-1}%
}\left( 1-\varkappa _{\mathsf{k}}\right) ^{\mathsf{n}^{\ast }\mathsf{-1-j}%
}\left( A-\varkappa _{\mathsf{k}}\right) ^{\mathsf{j}}\mathsf{q}_{\mathsf{k}}%
\mathsf{,\hat{p}}_{\beta ,\mathsf{Gibbs}}\right) _{2}
\end{equation*}%
to have $\mathsf{\hat{q}}_{\mathsf{k}}:=\mathsf{q}_{\mathsf{k}}+\gamma _{%
\mathsf{k}}\mathsf{\hat{p}}_{\beta ,\mathsf{Gibbs}}\in $ $\ker \left(
A-\varkappa _{\mathsf{k}}\right) ^{\mathsf{n}^{\ast }}$with $\mathsf{\hat{q}}%
_{\mathsf{k}}\neq 0$, hence $\mathsf{\hat{q}}_{\mathsf{k}}\in E_{\varkappa _{%
\mathsf{k}}}(A)$ according to Proposition 2. Arguing then as in the first
part of the proof and taking (\ref{inclusionter}) into account we obtain%
\begin{equation*}
\dbigvee\limits_{\mathsf{k=1}}^{+\infty }R_{\varkappa _{\mathsf{k}%
}}(A+P)\subseteq \dbigvee\limits_{\mathsf{k=1}}^{+\infty }E_{\nu _{\mathsf{k}%
}}(A).\text{ \ \ \ }\blacksquare
\end{equation*}

\bigskip

The preceding considerations now lead to the following result:

\bigskip

\textbf{Theorem 2.} \textit{Let us assume that }$\lambda _{\mathsf{m+1}%
}>\lambda _{\mathsf{m}}$ \textit{for every} $\mathsf{m}\in \mathbb{N}^{+}$%
\textit{, and let us impose the additional restriction }$\alpha \in \left(
1,3\right) $\textit{\ on the parameter introduced in (\ref{suitableconstants}%
). Then the set of all }$\mathsf{\hat{p}}_{\mathsf{k}}$ \textit{is complete
in }$l_{\mathbb{C}}^{2}$\textit{, that is,}%
\begin{equation}
l_{\mathbb{C}}^{2}=\dbigvee\limits_{\mathsf{k=1}}^{+\infty }E_{\nu _{\mathsf{%
k}}}(A).  \label{completeness}
\end{equation}

\bigskip

\textbf{Proof. }According to (\ref{spaces}) it is sufficient to show that
the set of all root vectors of the operator $A+P$ is complete in $l_{\mathbb{%
C}}^{2}$. To this end we realize $A+P$ as a perturbation of a linear,
bounded, invertible self-adjoint operator $H$ of finite order in the sense
of Theorem 8.1 in Chapter V of \cite{gohbergkrein}, namely, 
\begin{equation}
A+P=H\left( I+S+H^{-1}P\right)  \label{perturbation}
\end{equation}%
where $I$ stands for the identity operator in $l_{\mathbb{C}}^{2}$, $S$ for
a linear compact operator and $H^{-1}P$ compact. In order to achieve that we
define $H$ and $S$ by their matrix elements%
\begin{equation}
h_{\mathsf{m,n}}:=-b_{\mathsf{m}}\delta _{\mathsf{m,n}}
\label{firstoperator}
\end{equation}%
and%
\begin{equation}
s_{\mathsf{m,n}}:=-\frac{r_{\mathsf{m,n}}}{b_{\mathsf{m}}}
\label{secondoperator}
\end{equation}%
for all $\mathsf{m,n}\in \mathbb{N}^{+}$, respectively, where $r_{\mathsf{m,n%
}}$ is given by (\ref{transitionrate}) and $b_{\mathsf{m}}$ by (\ref%
{shorthand}). It is easily verified from (\ref{firstoperator}) that $H$ is
trace-class and thereby of finite order, the other required properties of $H$
being obvious. Using arguments similar to those invoked in the proof of
Proposition 1, it is equally straightforward to check that $S$ is also
trace-class and thereby compact. As for $H^{-1}P$ we have%
\begin{equation*}
H^{-1}P\mathsf{q=}\left( \mathsf{q,\hat{p}}_{\beta ,\mathsf{Gibbs}}\right)
_{2}H^{-1}\mathsf{\hat{p}}_{\beta ,\mathsf{Gibbs}}
\end{equation*}%
for every $\mathsf{q}\in l_{\mathbb{C}}^{2}$, where%
\begin{equation*}
\left( H^{-1}\mathsf{\hat{p}}_{\beta ,\mathsf{Gibbs}}\right) _{\mathsf{m}}=-%
\frac{\hat{p}_{\beta ,\mathsf{Gibbs,m}}}{b_{\mathsf{m}}}=c_{\alpha ,\beta
}\exp \left[ -\frac{3-\alpha }{2}\beta \lambda _{\mathsf{m}}\right]
\end{equation*}%
for each $\mathsf{m}\in \mathbb{N}^{+}$ and some irrelevant constant $%
c_{\alpha ,\beta }\in \mathbb{R}$. But we have assumed $\alpha \in \left(
1,3\right) $ and therefore, changing the value of $c_{\alpha ,\beta }$
whenever necessary, we obtain 
\begin{equation*}
H^{-1}\mathsf{\hat{p}}_{\beta ,\mathsf{Gibbs}}=c_{\alpha ,\beta }\mathsf{%
\hat{p}}_{\frac{3-\alpha }{2}\beta ,\mathsf{Gibbs}}
\end{equation*}%
so that $H^{-1}P$ turns out to be a bounded operator of rank one in $l_{%
\mathbb{C}}^{2}$ and thereby also compact. Finally, using (\ref{matrix}), (%
\ref{firstoperator}) and (\ref{secondoperator}) it is easily verified that
the relation%
\begin{equation*}
A=H\left( I+S\right)
\end{equation*}%
is valid, which is equivalent to (\ref{perturbation}). Since (\ref{kernelbis}%
) holds we may therefore apply Theorem 8.1 in Chapter V of \cite%
{gohbergkrein} to conclude that%
\begin{equation*}
l_{\mathbb{C}}^{2}=\dbigvee\limits_{\mathsf{k=1}}^{+\infty }R_{\varkappa _{%
\mathsf{k}}}(A+P).\text{ \ \ \ }\blacksquare
\end{equation*}

\bigskip

Provided we impose an additional restriction on the sequence $\left( \lambda
_{\mathsf{m}}\right) _{\mathsf{m}\in \mathbb{N}^{+}}$, whose r\^{o}le is to
control the gap between any two successive elements, we now proceed by
showing that the set of all $\mathsf{\hat{p}}_{\mathsf{k}}$ actually
constitute a basis of $l_{\mathbb{C}}^{2}$. In a Hilbert space setting this
means that there exists a unique sequence $\left( \mathsf{\hat{q}}_{\mathsf{k%
}}\right) _{\mathsf{k}\in \mathbb{N}^{+}}$ biorthogonal to $\left( \mathsf{%
\hat{p}}_{\mathsf{k}}\right) _{\mathsf{k}\in \mathbb{N}^{+}}$\ such that
every $\mathsf{p}\in $\ $l_{\mathbb{C}}^{2}$\ may be expanded in a unique
way as the norm-convergent series%
\begin{equation}
\mathsf{p}\text{ }=\dsum\limits_{\mathsf{k=1}}^{+\infty }\left( \mathsf{p,%
\hat{q}}_{\mathsf{k}}\right) _{2}\mathsf{\hat{p}}_{\mathsf{k}}  \label{basis}
\end{equation}%
(see, e.g., Chapter VI in \cite{gohbergkrein}). This, in turn, will lead to
the following spectral result:

\bigskip

\textbf{Main Theorem. }\textit{With }$\alpha \in \left( 1,3\right) $ \textit{%
and} $\theta \in \left( 0,\frac{3-\alpha }{2}\beta \right) $, \textit{let us
assume that} 
\begin{equation}
\lambda _{\mathsf{m+1}}-\lambda _{\mathsf{m}}\geqslant c\exp \left[ -\theta
\lambda _{\mathsf{m}}\right]  \label{gapcontrol}
\end{equation}%
\textit{for every} $\mathsf{m}\in \mathbb{N}^{+}$ \textit{and} \textit{some} 
$c>0$ \textit{independent of }$\mathsf{m}$. \textit{Then the set }$\left( 
\mathsf{\hat{p}}_{\mathsf{k}}\right) _{\mathsf{k}\in \mathbb{N}^{+}}$\textit{%
\ of eigenvectors given by (\ref{generator}) provides a basis for }$l_{%
\mathbb{C}}^{2}$\textit{\ in the sense of (\ref{basis}). Moreover, for each }%
$\mathsf{p}\in $\textit{\ }$l_{\mathbb{C}}^{2}$ \textit{and every }$\tau \in %
\left[ 0,+\infty \right) $ \textit{we have the norm-convergent spectral
decomposition}%
\begin{equation}
\exp \left[ \tau A\right] \mathsf{p}=\dsum\limits_{\mathsf{k=1}}^{+\infty
}\left( \mathsf{p,\hat{q}}_{\mathsf{k}}\right) _{2}\exp \left[ \tau \nu _{%
\mathsf{k}}\right] \mathsf{\hat{p}}_{\mathsf{k}}  \label{semigrouprep}
\end{equation}%
\textit{of the semigroup }$\exp \left[ \tau A\right] _{\tau \in \left[
0,+\infty \right) }$\textit{\ generated by }$A.$

\bigskip

The proof of this theorem will be somewhat indirect and rests upon several
preparatory results. We begin with the description of the biorthogonal
sequence we alluded to above, and refer the reader to Appendix B for an
alternative construction:

\bigskip

\textbf{Proposition 4.} \textit{If} $\alpha \in \left( 1,3\right) $, \textit{%
there exists a unique sequence} $\left( \mathsf{\hat{q}}_{\mathsf{k}}\right)
_{\mathsf{k}\in \mathbb{N}^{+}}$ \textit{biorthogonal to} $\left( \mathsf{%
\hat{p}}_{\mathsf{k}}\right) _{\mathsf{k}\in \mathbb{N}^{+}}$.

\bigskip

\textbf{Proof. }Let us consider the orthogonal projection%
\begin{equation*}
Q:l_{\mathbb{C}}^{2}\mapsto \left( \dbigvee\limits_{\mathsf{k=2}}^{+\infty
}E_{\nu _{\mathsf{k}}}(A)\right) ^{\perp }.
\end{equation*}%
We then have $\left\Vert Q\mathsf{\hat{p}}_{1}\right\Vert _{2}^{2}=\left( 
\mathsf{\hat{p}}_{1},Q\mathsf{\hat{p}}_{1}\right) _{2}\neq 0$ as a
consequence of a general fact proved in Appendix B, so that we may define%
\begin{equation}
\mathsf{\hat{q}}_{\mathsf{1}}:=\left\Vert Q\mathsf{\hat{p}}_{1}\right\Vert
_{2}^{-2}Q\mathsf{\hat{p}}_{1}.  \label{firstelement}
\end{equation}%
Consequently we get%
\begin{equation}
\left( \mathsf{\hat{p}}_{\mathsf{j}},\mathsf{\hat{q}}_{\mathsf{1}}\right)
_{2}=\delta _{\mathsf{j,1}}  \label{biorthogonality}
\end{equation}%
for every $\mathsf{j}\in \mathbb{N}^{+}$. Furthermore, since $A$ and $%
A^{\ast }$ are compact and the $\nu _{\mathsf{k}}$ are real, the eigenvalue
equations for them when $\nu _{\mathsf{k}}\neq 0$ read $\left( A-\nu _{%
\mathsf{k}}\right) \mathsf{\hat{p}}_{\mathsf{k}}=0$ and $\left( A^{\ast
}-\nu _{\mathsf{k}}\right) \mathsf{\hat{q}}_{\mathsf{k}}=0$ respectively,
with $\mathsf{\hat{q}}_{\mathsf{k}}\in l_{\mathbb{C}}^{2}$ given by (\ref%
{generatorbis}) in the second case. Therefore we have%
\begin{equation*}
\left( \nu _{\mathsf{j}}-\nu _{\mathsf{k}}\right) \left( \mathsf{\hat{p}}_{%
\mathsf{j}}\mathsf{,\hat{q}}_{\mathsf{k}}\right) _{2}=\left( A\mathsf{\hat{p}%
}_{\mathsf{j}}\mathsf{,\hat{q}}_{\mathsf{k}}\right) _{2}-\left( \mathsf{\hat{%
p}}_{\mathsf{j}}\mathsf{,A^{\ast }\hat{q}}_{\mathsf{k}}\right) _{2}=0,
\end{equation*}%
so that if $\mathsf{j,k}\geqslant 2$ with $\mathsf{j\neq k}$ then $\nu _{%
\mathsf{j}}\neq \nu _{\mathsf{k}}$ and so%
\begin{equation}
\left( \mathsf{\hat{p}}_{\mathsf{j}}\mathsf{,\hat{q}}_{\mathsf{k}}\right)
_{2}=0.  \label{biorthogonalitybis}
\end{equation}%
A similar argument shows that 
\begin{equation*}
\left( \mathsf{\hat{p}}_{1}\mathsf{,\hat{q}}_{\mathsf{k}}\right) _{2}=0
\end{equation*}%
\textsf{for }$\mathsf{k}\geqslant 2$ and moreover we can impose $\left( 
\mathsf{\hat{p}}_{\mathsf{k}}\mathsf{,\hat{q}}_{\mathsf{k}}\right) _{2}=1$
by normalizing the eigenvectors accordingly, remembering that we may not
have $\left( \mathsf{\hat{p}}_{\mathsf{k}}\mathsf{,\hat{q}}_{\mathsf{k}%
}\right) _{2}=0$ since this and (\ref{biorthogonality}) for $\mathsf{j\neq k}
$ would imply $\mathsf{\hat{q}}_{\mathsf{k}}=0$ by virtue of (\ref%
{completeness}). Altogether we have%
\begin{equation*}
\left( \mathsf{\hat{p}}_{\mathsf{j}}\mathsf{,\hat{q}}_{\mathsf{k}}\right)
_{2}=\delta _{\mathsf{j,k}}
\end{equation*}%
for all $\mathsf{j,k}\in \mathbb{N}^{+}$ as required, and the uniqueness of
such a sequence is an immediate consequence of the completeness of the $%
\mathsf{\hat{p}}_{\mathsf{k}}$ guaranteed by Theorem 2. \ \ $\blacksquare $

\bigskip

Our next step consists in renormalizing the $\mathsf{\hat{p}}_{\mathsf{k}}$
by defining the $\mathsf{\hat{r}}_{\mathsf{k}}$ as%
\begin{equation}
\mathsf{\hat{r}}_{\mathsf{k}}:=\left\{ 
\begin{array}{c}
\mathsf{\hat{p}}_{1}\text{ \ \ \ for }\mathsf{k=1,} \\ 
\\ 
\mathsf{\hat{p}}_{\mathsf{k}}-\mathsf{\hat{p}}_{1},\text{ \ \ \ for }\mathsf{%
k}\in \left\{ 2,3,...\right\} ,%
\end{array}%
\right.  \label{renormalization}
\end{equation}%
and in proving that the sequence $\left( \mathsf{\hat{r}}_{\mathsf{k}%
}\right) _{\mathsf{k}\in \mathbb{N}^{+}}$ constitutes a basis of $l_{\mathbb{%
C}}^{2}$. It first follows from (\ref{shorthand}) and (\ref{generator}) that%
\begin{equation}
\mathsf{\hat{r}}_{\mathsf{k,m}}=-\frac{\nu _{\mathsf{k}}\exp \left[ -\beta
\lambda _{\mathsf{m}}\right] }{Z_{\frac{\alpha +1}{2}\beta }\left( \nu _{%
\mathsf{k}}+b_{\mathsf{m}}\right) }  \label{generatorquarto}
\end{equation}%
for each $\mathsf{k\in }\left\{ 2,3,...\right\} $ and every $\mathsf{m}\in 
\mathbb{N}^{+}$, and from the proof of Proposition 4 that the unique
sequence $\left( \mathsf{\hat{s}}_{\mathsf{k}}\right) _{\mathsf{k}\in 
\mathbb{N}^{+}}$ biorthogonal to $\left( \mathsf{\hat{r}}_{\mathsf{k}%
}\right) _{\mathsf{k}\in \mathbb{N}^{+}}$ is given by%
\begin{equation}
\mathsf{\hat{s}}_{\mathsf{k}}=\left\{ 
\begin{array}{c}
\left\Vert \tilde{Q}\mathsf{\hat{p}}_{\mathsf{1}}\right\Vert _{2}^{-2}\tilde{%
Q}\mathsf{\hat{p}}_{\mathsf{1}}\text{ \ \ for }\mathsf{k=1,} \\ 
\\ 
\mathsf{\hat{q}}_{\mathsf{k}}\text{\ \ for }\mathsf{k}\in \left\{
2,3,...\right\} ,%
\end{array}%
\right.  \label{biorthosequence}
\end{equation}%
where%
\begin{equation*}
\tilde{Q}:l_{\mathbb{C}}^{2}\mapsto \left( \func{cl}\func{span}\left\{ 
\mathsf{\hat{r}}_{\mathsf{k}},\text{ }\mathsf{k}\in \left\{ 2,3,...\right\}
\right\} \right) ^{\perp }
\end{equation*}%
denotes the orthogonal projection onto the indicated space, with $\func{cl}%
\func{span}\mathsf{\ }$referring to the closure of the span in $l_{\mathbb{C}%
}^{2}$. In order to prove that the $\mathsf{\hat{r}}_{\mathsf{k}}$ provide a
basis for $l_{\mathbb{C}}^{2}$ we first need some estimates related to the
localization properties of the $\nu _{\mathsf{k}}$. We begin with the
following:

\bigskip

\textbf{Lemma 1. }\textit{Under the sole conditions }$\alpha >1$, $\beta >0$%
\textit{, there exists a constant }$c_{\alpha ,\beta }>0$\textit{\ depending
on }$\alpha $ \textit{and} $\beta $\textit{\ such that the estimate}%
\begin{equation}
\mathsf{b}_{\mathsf{m}}-\mathsf{b}_{\mathsf{m+1}}\geqslant c_{\alpha ,\beta
}\exp \left[ -\left( \frac{\alpha -1}{2}\beta +\theta \right) \lambda _{%
\mathsf{m}}\right]  \label{inequality4}
\end{equation}%
\textit{holds for each }$\mathsf{m}\in \mathbb{N}^{+}$ \textit{and every} $%
\theta >0$\textit{, where }$\mathsf{b}_{\mathsf{m}}$\textit{\ is given by (%
\ref{shorthand}).}

\bigskip

\textbf{Proof. }From (\ref{shorthand}) we first have%
\begin{eqnarray}
&&\mathsf{b}_{\mathsf{m}}-\mathsf{b}_{\mathsf{m+1}}  \notag \\
&=&Z_{\frac{\alpha +1}{2}\beta }\exp \left[ -\frac{\alpha -1}{2}\beta
\lambda _{\mathsf{m}}\right] \left( 1-\exp \left[ -\frac{\alpha -1}{2}\beta
\left( \lambda _{\mathsf{m+1}}-\lambda _{\mathsf{m}}\right) \right] \right)
\label{inequality3} \\
&\geqslant &Z_{\frac{\alpha +1}{2}\beta }\exp \left[ -\frac{\alpha -1}{2}%
\beta \lambda _{\mathsf{m}}\right] \left( 1+\frac{2}{\left( \alpha -1\right)
\beta }\left( \lambda _{\mathsf{m+1}}-\lambda _{\mathsf{m}}\right)
^{-1}\right) ^{-1}  \notag
\end{eqnarray}%
as a consequence of the elementary inequality%
\begin{equation*}
1-\exp \left[ -x\right] \geqslant \left( 1+x^{-1}\right) ^{-1}
\end{equation*}%
valid for every $x>0$. Furthermore, from (\ref{gapcontrol}) we get the lower
bounds 
\begin{eqnarray*}
&&\left( 1+\frac{2}{\left( \alpha -1\right) \beta }\left( \lambda _{\mathsf{%
m+1}}-\lambda _{\mathsf{m}}\right) ^{-1}\right) ^{-1} \\
&\geqslant &\left( 1+\frac{2}{c\left( \alpha -1\right) \beta }\exp \left[
\theta \lambda _{\mathsf{m}}\right] \right) ^{-1}=\exp \left[ -\theta
\lambda _{\mathsf{m}}\right] \left( \exp \left[ -\theta \lambda _{\mathsf{m}}%
\right] +\frac{2}{c\left( \alpha -1\right) \beta }\right) ^{-1} \\
&\geqslant &\exp \left[ -\theta \lambda _{\mathsf{m}}\right] \left( 1+\frac{2%
}{c\left( \alpha -1\right) \beta }\right) ^{-1}
\end{eqnarray*}%
for the third factor on the right-hand side of the inequality in (\ref%
{inequality3}) since $\theta >0$, and since we may assume $\lambda _{\mathsf{%
m}}>0$ for each $\mathsf{m}$ without restricting the generality. The
substitution of the last estimate into (\ref{inequality3}) then leads to (%
\ref{inequality4}) with an obvious choice for $c_{\alpha ,\beta }$. \ \ $%
\blacksquare $

\bigskip

Next we have:

\bigskip

\textbf{Lemma 2. }\textit{Under the sole conditions }$\alpha >1$, $\beta >0$,%
\textit{\ there exists a constant }$\hat{c}_{\alpha ,\beta }>0$ \textit{%
depending on }$\alpha $\textit{\ and }$\beta $ \textit{such that the
inequality}%
\begin{equation}
\left\vert \nu _{\mathsf{k}}+\mathsf{b}_{\mathsf{m}}\right\vert \geqslant
c_{\alpha ,\beta }\exp \left[ -\left( \frac{\alpha -1}{2}\beta +\theta
\right) \lambda _{\mathsf{m}}\right]  \label{inequality1}
\end{equation}%
\textit{holds for all sufficiently large }$\mathsf{k}$\textit{, each }$%
\mathsf{m}\neq \mathsf{k}$ \textit{and every} $0<\theta <\beta $\textsf{. }%
\textit{Moreover, for all sufficiently large }$\mathsf{k}$ \textit{and} $%
\mathsf{m}=\mathsf{k}$ \textit{we have}%
\begin{equation}
\left\vert \nu _{\mathsf{k}}+\mathsf{b}_{\mathsf{k}}\right\vert \leq \hat{c}%
_{\alpha ,\beta ,\lambda _{1}}\exp \left[ -\frac{\alpha +1}{2}\beta \lambda
_{\mathsf{k}}\right]  \label{inequality7}
\end{equation}%
\textit{for some suitable} $\hat{c}_{\alpha ,\beta ,\lambda _{1}}>0$.

\bigskip

\textbf{Proof.} Let us first consider the case $\mathsf{m}>\mathsf{k}\in
\left\{ 2,3,...\right\} $. From the localization property of the eigenvalues
stated in (c) of Theorem 1 we then have%
\begin{equation}
\mathsf{b}_{\mathsf{m}}-\mathsf{b}_{\mathsf{k-1}}<\nu _{\mathsf{k}}+\mathsf{b%
}_{\mathsf{m}}<\mathsf{b}_{\mathsf{m}}-\mathsf{b}_{\mathsf{k}}<0
\label{localization}
\end{equation}%
and thereby%
\begin{equation*}
\left\vert \nu _{\mathsf{k}}+\mathsf{b}_{\mathsf{m}}\right\vert \geqslant 
\mathsf{b}_{\mathsf{k}}-\mathsf{b}_{\mathsf{m}}\geqslant \mathsf{b}_{\mathsf{%
m-1}}-\mathsf{b}_{\mathsf{m}}
\end{equation*}%
since $\mathsf{m-1\geqslant k}$. Therefore, applying (\ref{inequality3})
with $\mathsf{m-1}$ instead of $\mathsf{m}$ and using the fact that $\lambda
_{\mathsf{m}}>\lambda _{\mathsf{m-1}}$ we obtain 
\begin{equation}
\left\vert \nu _{\mathsf{k}}+\mathsf{b}_{\mathsf{m}}\right\vert \geqslant
c_{\alpha ,\beta }\exp \left[ -\left( \frac{\alpha -1}{2}\beta +\theta
\right) \lambda _{\mathsf{m-1}}\right] \geqslant c_{\alpha ,\beta }\exp %
\left[ -\left( \frac{\alpha -1}{2}\beta +\theta \right) \lambda _{\mathsf{m}}%
\right] ,  \label{inequality2}
\end{equation}%
which is (\ref{inequality1}) for this case.

Let us now assume that $\mathsf{m}<\mathsf{k-1}$. From (\ref{localization})
we have this time%
\begin{equation*}
\left\vert \nu _{\mathsf{k}}+\mathsf{b}_{\mathsf{m}}\right\vert \geqslant 
\mathsf{b}_{\mathsf{m}}-\mathsf{b}_{\mathsf{k-1}}\geqslant \mathsf{b}_{%
\mathsf{m}}-\mathsf{b}_{\mathsf{m+1}}
\end{equation*}%
since $\mathsf{m+1}\leqslant \mathsf{k-1}$, so that (\ref{inequality1})
again follows directly from (\ref{inequality3}).

The remaining cases are a bit trickier. Thus, let us take $\mathsf{m}=%
\mathsf{k}$ and observe that (\ref{characterizationter}) along with the
localization property of the eigenvalues imply the identity%
\begin{equation}
\frac{\exp \left[ -\frac{\alpha +1}{2}\beta \lambda _{\mathsf{k}}\right] }{%
\left\vert \nu _{\mathsf{k}}+b_{\mathsf{k}}\right\vert }=\dsum\limits_{%
\mathsf{m=1}}^{\mathsf{k-1}}\frac{\exp \left[ -\frac{\alpha +1}{2}\beta
\lambda _{\mathsf{m}}\right] }{\left\vert \nu _{\mathsf{k}}+b_{\mathsf{m}%
}\right\vert }-\dsum\limits_{\mathsf{m=k+1}}^{\mathsf{+\infty }}\frac{\exp %
\left[ -\frac{\alpha +1}{2}\beta \lambda _{\mathsf{m}}\right] }{\left\vert
\nu _{\mathsf{k}}+b_{\mathsf{m}}\right\vert }  \label{identity}
\end{equation}%
for every $\mathsf{k}\in \left\{ 2,3,...\right\} $. We then proceed by
getting a lower bound for each term on the right-hand side of (\ref{identity}%
). On the one hand we have%
\begin{equation}
\dsum\limits_{\mathsf{m=1}}^{\mathsf{k-1}}\frac{\exp \left[ -\frac{\alpha +1%
}{2}\beta \lambda _{\mathsf{m}}\right] }{\left\vert \nu _{\mathsf{k}}+b_{%
\mathsf{m}}\right\vert }\geqslant \frac{\exp \left[ -\frac{\alpha +1}{2}%
\beta \lambda _{\mathsf{1}}\right] }{\left\vert \nu _{\mathsf{k}}+b_{\mathsf{%
1}}\right\vert }\geqslant \frac{\exp \left[ -\frac{\alpha +1}{2}\beta
\lambda _{\mathsf{1}}\right] }{b_{\mathsf{1}}}=\frac{\exp \left[ -\beta
\lambda _{\mathsf{1}}\right] }{Z_{\frac{\alpha +1}{2}\beta }}
\label{inequality5}
\end{equation}%
as a consequence of (\ref{shorthand}) since $\left\vert \nu _{\mathsf{k}}+b_{%
\mathsf{1}}\right\vert \leqslant b_{\mathsf{1}}$. On the other hand we obtain%
\begin{equation}
\dsum\limits_{\mathsf{m=k+1}}^{\mathsf{+\infty }}\frac{\exp \left[ -\frac{%
\alpha +1}{2}\beta \lambda _{\mathsf{m}}\right] }{\left\vert \nu _{\mathsf{k}%
}+b_{\mathsf{m}}\right\vert }\leqslant \frac{1}{c_{\alpha ,\beta }}%
\dsum\limits_{\mathsf{m=k+1}}^{\mathsf{+\infty }}\exp \left[ -\left( \beta
-\theta \right) \lambda _{\mathsf{m}}\right]  \label{inequality6}
\end{equation}%
from (\ref{inequality2}), so that the substitution of (\ref{inequality5})
and (\ref{inequality6}) into (\ref{identity}) leads to%
\begin{equation}
\frac{\exp \left[ -\frac{\alpha +1}{2}\beta \lambda _{\mathsf{k}}\right] }{%
\left\vert \nu _{\mathsf{k}}+b_{\mathsf{k}}\right\vert }\geqslant \frac{\exp %
\left[ -\beta \lambda _{\mathsf{1}}\right] }{Z_{\frac{\alpha +1}{2}\beta }}-%
\frac{1}{c_{\alpha ,\beta }}\dsum\limits_{\mathsf{m=k+1}}^{\mathsf{+\infty }%
}\exp \left[ -\left( \beta -\theta \right) \lambda _{\mathsf{m}}\right] .
\label{inequality8}
\end{equation}%
Since $\theta <\beta $, the second term on the right-hand side of the
preceding expression tends to zero as $\mathsf{k}$ becomes large by virtue
of (\ref{partitionfunction}). In particular, there exists $\mathsf{k}^{\ast
} $ such that for every $\mathsf{k\geqslant k}^{\ast }$ we have 
\begin{equation*}
\dsum\limits_{\mathsf{m=k+1}}^{\mathsf{+\infty }}\exp \left[ -\left( \beta
-\theta \right) \lambda _{\mathsf{m}}\right] \leqslant \frac{c_{\alpha
,\beta }}{2Z_{\frac{\alpha +1}{2}\beta }}\exp \left[ -\beta \lambda _{%
\mathsf{1}}\right] ,
\end{equation*}%
which gives%
\begin{equation*}
\frac{\exp \left[ -\frac{\alpha +1}{2}\beta \lambda _{\mathsf{k}}\right] }{%
\left\vert \nu _{\mathsf{k}}+b_{\mathsf{k}}\right\vert }\geqslant \frac{\exp %
\left[ -\beta \lambda _{\mathsf{1}}\right] }{2Z_{\frac{\alpha +1}{2}\beta }}
\end{equation*}%
according to (\ref{inequality8}), and thereby (\ref{inequality7}) with an
obvious choice for $\hat{c}_{\alpha ,\beta ,\lambda _{1}}$.

It remains to consider the case $\mathsf{m}=\mathsf{k-1}$. Using once again
the localization property we have%
\begin{equation*}
\left\vert \nu _{\mathsf{k}}+b_{\mathsf{k-1}}\right\vert =\mathsf{b}_{%
\mathsf{k-1}}-\mathsf{b}_{\mathsf{k}}-\left\vert \nu _{\mathsf{k}}+b_{%
\mathsf{k}}\right\vert ,
\end{equation*}%
and therefore the estimate 
\begin{eqnarray}
\left\vert \nu _{\mathsf{k}}+b_{\mathsf{k-1}}\right\vert &\geqq &c_{\alpha
,\beta }\exp \left[ -\left( \frac{\alpha -1}{2}\beta +\theta \right) \lambda
_{\mathsf{k-1}}\right] -\hat{c}\exp \left[ -\frac{\alpha +1}{2}\beta \lambda
_{\mathsf{k-1}}\right]  \label{inequality9} \\
&=&\exp \left[ -\left( \frac{\alpha -1}{2}\beta +\theta \right) \lambda _{%
\mathsf{k-1}}\right] \left\{ c_{\alpha ,\beta }-\hat{c}\exp \left[ -\left(
\beta -\theta \right) \lambda _{\mathsf{k-1}}\right] \right\}  \notag
\end{eqnarray}%
for every $\mathsf{k\geqslant k}^{\ast }$ as a consequence of Lemma 1 with $%
\mathsf{m}=\mathsf{k-1}$ and (\ref{inequality7}). Now since $\theta <\beta $
and $\lambda _{\mathsf{k}}\rightarrow +\infty $ as $\mathsf{k\rightarrow
+\infty }$, there exists $\mathsf{k}^{\ast \ast }$ such that for every $%
\mathsf{k\geqslant k}^{\ast \ast }$ we have%
\begin{equation*}
\hat{c}\exp \left[ -\left( \beta -\theta \right) \lambda _{\mathsf{k-1}}%
\right] \leqslant \frac{c_{\alpha ,\beta }}{2}.
\end{equation*}%
Consequently, the substitution of the preceding relation into (\ref%
{inequality9}) gives%
\begin{equation*}
\left\vert \nu _{\mathsf{k}}+b_{\mathsf{k-1}}\right\vert \geqslant \frac{%
c_{\alpha ,\beta }}{2}\exp \left[ -\left( \frac{\alpha -1}{2}\beta +\theta
\right) \lambda _{\mathsf{k-1}}\right]
\end{equation*}%
for all sufficiently large $\mathsf{k}$. \ \ $\blacksquare $

\bigskip

The critical result is now the following:

\bigskip

\textbf{Proposition 5.} \textit{Let us assume that the same hypotheses as in
the main theorem hold. Then for every }$\mathsf{p}\in l_{\mathbb{C}}^{2}$%
\textit{\ we have the norm-convergent series expansion}%
\begin{equation}
\mathsf{p}\text{ }=\dsum\limits_{\mathsf{k=1}}^{+\infty }\left( \mathsf{p,}%
\left\Vert \mathsf{\hat{r}}_{\mathsf{k}}\right\Vert _{2}\mathsf{\hat{s}}_{%
\mathsf{k}}\right) _{2}\frac{\mathsf{\hat{r}}_{\mathsf{k}}}{\left\Vert 
\mathsf{\hat{r}}_{\mathsf{k}}\right\Vert _{2}}  \label{basisbis}
\end{equation}%
\textit{where} $\mathsf{\hat{r}}_{\mathsf{k}}$ \textit{and} $\mathsf{\hat{s}}%
_{\mathsf{k}}$ \textit{are given by (\ref{renormalization}) and (\ref%
{biorthosequence}), respectively}. \textit{Equivalently, the sequence} $%
\left( \mathsf{\hat{r}}_{\mathsf{k}}\right) _{\mathsf{k}\in \mathbb{N}^{+}}$%
\textit{\ provides a basis for }$l_{\mathbb{C}}^{2}$.

\bigskip

\textbf{Proof.} It is plain that the $\mathsf{\hat{r}}_{\mathsf{k}}$ form a
complete set in $l_{\mathbb{C}}^{2}$ since the $\mathsf{\hat{p}}_{\mathsf{k}%
} $ do. Next, we prove that%
\begin{equation}
\sum_{\mathsf{j=1}}^{+\infty }\sum_{\mathsf{k=1},\mathsf{k\neq j}}^{+\infty
}\left\vert \left( \frac{\mathsf{\hat{r}}_{\mathsf{j}}}{\left\Vert \mathsf{%
\hat{r}}_{\mathsf{j}}\right\Vert _{2}},\frac{\mathsf{\hat{r}}_{\mathsf{k}}}{%
\left\Vert \mathsf{\hat{r}}_{\mathsf{k}}\right\Vert _{2}}\right)
_{2}\right\vert ^{2}<+\infty  \label{convergence}
\end{equation}%
by using the inequalities we provided in Lemma 2. For the inner product in
the preceding expression we first have%
\begin{equation*}
\left( \frac{\mathsf{\hat{r}}_{\mathsf{j}}}{\left\Vert \mathsf{\hat{r}}_{%
\mathsf{j}}\right\Vert _{2}},\frac{\mathsf{\hat{r}}_{\mathsf{k}}}{\left\Vert 
\mathsf{\hat{r}}_{\mathsf{k}}\right\Vert _{2}}\right) _{2}=\frac{1}{%
\left\Vert \mathsf{\hat{r}}_{\mathsf{j}}\right\Vert _{2}\left\Vert \mathsf{%
\hat{r}}_{\mathsf{k}}\right\Vert _{2}}\left( \mathsf{\hat{r}}_{\mathsf{j,j}}%
\mathsf{\hat{r}}_{\mathsf{k,j}}+\mathsf{\hat{r}}_{\mathsf{j,k}}\mathsf{\hat{r%
}}_{\mathsf{k,k}}+\dsum\limits_{\mathsf{m}=1,\mathsf{m\neq j,k}}^{+\infty }%
\mathsf{\hat{r}}_{\mathsf{j,m}}\mathsf{\hat{r}}_{\mathsf{k,m}}\right)
\end{equation*}%
and therefore the estimate%
\begin{equation}
\left\vert \left( \frac{\mathsf{\hat{r}}_{\mathsf{j}}}{\left\Vert \mathsf{%
\hat{r}}_{\mathsf{j}}\right\Vert _{2}},\frac{\mathsf{\hat{r}}_{\mathsf{k}}}{%
\left\Vert \mathsf{\hat{r}}_{\mathsf{k}}\right\Vert _{2}}\right)
_{2}\right\vert \leqslant \frac{\left\vert \mathsf{\hat{r}}_{\mathsf{k,j}%
}\right\vert }{\left\vert \mathsf{\hat{r}}_{\mathsf{k,k}}\right\vert }+\frac{%
\left\vert \mathsf{\hat{r}}_{\mathsf{j,k}}\right\vert }{\left\vert \mathsf{%
\hat{r}}_{\mathsf{j,j}}\right\vert }+\dsum\limits_{\mathsf{m}=1,\mathsf{%
m\neq j,k}}^{+\infty }\frac{\left\vert \mathsf{\hat{r}}_{\mathsf{j,m}%
}\right\vert \left\vert \mathsf{\hat{r}}_{\mathsf{k,m}}\right\vert }{%
\left\vert \mathsf{\hat{r}}_{\mathsf{j,j}}\right\vert \left\vert \mathsf{%
\hat{r}}_{\mathsf{k,k}}\right\vert }  \label{inequality10}
\end{equation}%
by virtue of the inequalities $\left\Vert \mathsf{\hat{r}}_{\mathsf{j}%
}\right\Vert _{2}\geqslant \left\vert \mathsf{\hat{r}}_{\mathsf{j,j}%
}\right\vert $ and $\left\Vert \mathsf{\hat{r}}_{\mathsf{k}}\right\Vert
_{2}\geqslant \left\vert \mathsf{\hat{r}}_{\mathsf{k,k}}\right\vert $. Now
from (\ref{generatorquarto}) we have%
\begin{equation*}
\frac{\left\vert \mathsf{\hat{r}}_{\mathsf{k,j}}\right\vert }{\left\vert 
\mathsf{\hat{r}}_{\mathsf{k,k}}\right\vert }=\frac{\left\vert \nu _{\mathsf{k%
}}+b_{\mathsf{k}}\right\vert }{\left\vert \nu _{\mathsf{k}}+b_{\mathsf{j}%
}\right\vert }\exp \left[ -\beta \left( \lambda _{\mathsf{j}}-\lambda _{%
\mathsf{k}}\right) \right]
\end{equation*}%
and 
\begin{equation*}
\frac{\left\vert \mathsf{\hat{r}}_{\mathsf{j,k}}\right\vert }{\left\vert 
\mathsf{\hat{r}}_{\mathsf{j,j}}\right\vert }=\frac{\left\vert \nu _{\mathsf{j%
}}+b_{\mathsf{j}}\right\vert }{\left\vert \nu _{\mathsf{j}}+b_{\mathsf{k}%
}\right\vert }\exp \left[ -\beta \left( \lambda _{\mathsf{k}}-\lambda _{%
\mathsf{j}}\right) \right]
\end{equation*}%
for all $\mathsf{j}$ and $\mathsf{k}$, so that by means of inequalities (\ref%
{inequality1}) and (\ref{inequality7}) we obtain%
\begin{equation}
\frac{\left\vert \mathsf{\hat{r}}_{\mathsf{k,j}}\right\vert }{\left\vert 
\mathsf{\hat{r}}_{\mathsf{k,k}}\right\vert }\leqslant \frac{\hat{c}}{%
c_{\alpha ,\beta }}\exp \left[ -\left( \frac{3-\alpha }{2}\beta -\theta
\right) \lambda _{\mathsf{j}}\right] \exp \left[ -\frac{\alpha -1}{2}\beta
\lambda _{\mathsf{k}}\right]  \label{inequality11}
\end{equation}%
for all sufficiently large $\mathsf{k}$ and every $\mathsf{j\neq k}$, and
similarly%
\begin{equation}
\frac{\left\vert \mathsf{\hat{r}}_{\mathsf{j,k}}\right\vert }{\left\vert 
\mathsf{\hat{r}}_{\mathsf{j,j}}\right\vert }\leqslant \frac{\hat{c}}{%
c_{\alpha ,\beta }}\exp \left[ -\left( \frac{3-\alpha }{2}\beta -\theta
\right) \lambda _{\mathsf{k}}\right] \exp \left[ -\frac{\alpha -1}{2}\beta
\lambda _{\mathsf{j}}\right]  \label{inequality12}
\end{equation}%
for all sufficiently large $\mathsf{j}$ and every $\mathsf{k\neq j}$. By the
same token we have%
\begin{eqnarray}
&&\dsum\limits_{\mathsf{m}=1,\mathsf{m\neq j,k}}^{+\infty }\frac{\left\vert 
\mathsf{\hat{r}}_{\mathsf{j,m}}\right\vert \left\vert \mathsf{\hat{r}}_{%
\mathsf{k,m}}\right\vert }{\left\vert \mathsf{\hat{r}}_{\mathsf{j,j}%
}\right\vert \left\vert \mathsf{\hat{r}}_{\mathsf{k,k}}\right\vert }  \notag
\\
&\leqslant &\left( \frac{\hat{c}}{c_{\alpha ,\beta }}\right)
^{2}\dsum\limits_{\mathsf{m}=1,\mathsf{m\neq j,k}}^{+\infty }\exp \left[
-\left( \left( 3-\alpha \right) \beta -2\theta \right) \lambda _{\mathsf{m}}%
\right] \exp \left[ -\frac{\alpha -1}{2}\beta \left( \lambda _{\mathsf{j}%
}+\lambda _{\mathsf{k}}\right) \right]  \notag \\
&\leqslant &\left( \frac{\hat{c}}{c_{\alpha ,\beta }}\right) ^{2}Z_{\left(
3-\alpha \right) \beta -2\theta }\exp \left[ -\frac{\alpha -1}{2}\beta
\left( \lambda _{\mathsf{j}}+\lambda _{\mathsf{k}}\right) \right]
\label{inequality13}
\end{eqnarray}%
by virtue of (\ref{partitionfunction}) as a consequence of the hypotheses
regarding $\alpha $ and $\theta $, for all sufficiently large $\mathsf{j}$
and $\mathsf{k}$ with $\mathsf{j\neq k}$. Let us now define 
\begin{equation*}
\kappa :=\left( \frac{3-\alpha }{2}\beta -\theta \right) \wedge \frac{\alpha
-1}{2}\beta ,
\end{equation*}%
the smaller of the two numbers, which is positive. Then, by using estimates (%
\ref{inequality10})-(\ref{inequality13}) we get%
\begin{equation*}
\left\vert \left( \frac{\mathsf{\hat{r}}_{\mathsf{j}}}{\left\Vert \mathsf{%
\hat{r}}_{\mathsf{j}}\right\Vert _{2}},\frac{\mathsf{\hat{r}}_{\mathsf{k}}}{%
\left\Vert \mathsf{\hat{r}}_{\mathsf{k}}\right\Vert _{2}}\right)
_{2}\right\vert \leqslant \left( \frac{2\hat{c}}{c_{\alpha ,\beta }}+\left( 
\frac{\hat{c}}{c_{\alpha ,\beta }}\right) ^{2}Z_{\left( 3-\alpha \right)
\beta -2\theta }\right) \exp \left[ -\kappa \left( \lambda _{\mathsf{j}%
}+\lambda _{\mathsf{k}}\right) \right]
\end{equation*}%
and there exist $\mathsf{J},\mathsf{K}\in \mathbb{N}^{+}$ such that%
\begin{eqnarray*}
&&\sum_{\mathsf{j=J}}^{+\infty }\sum_{\mathsf{k=K},\mathsf{k\neq j}%
}^{+\infty }\left\vert \left( \frac{\mathsf{\hat{r}}_{\mathsf{j}}}{%
\left\Vert \mathsf{\hat{r}}_{\mathsf{j}}\right\Vert _{2}},\frac{\mathsf{\hat{%
r}}_{\mathsf{k}}}{\left\Vert \mathsf{\hat{r}}_{\mathsf{k}}\right\Vert _{2}}%
\right) _{2}\right\vert ^{2} \\
&\leqslant &c_{\alpha ,\beta ,\theta }\sum_{\mathsf{j=J}}^{+\infty }\sum_{%
\mathsf{k=K},\mathsf{k\neq j}}^{+\infty }\exp \left[ -2\kappa \left( \lambda
_{\mathsf{j}}+\lambda _{\mathsf{k}}\right) \right] \leqslant c_{\alpha
,\beta ,\theta }Z_{2\kappa }^{2}<\infty
\end{eqnarray*}%
for some $c_{\alpha ,\beta ,\theta }$ because of (\ref{partitionfunction}),
which proves (\ref{convergence}) and thus guarantees the existence of a
large enough $\mathsf{N}^{\ast }\in \mathbb{N}^{+}$ such that 
\begin{equation}
\sum_{\mathsf{j,k\geqslant N}^{\ast },\mathsf{j\neq k}}^{+\infty }\left\vert
\left( \frac{\mathsf{\hat{r}}_{\mathsf{j}}}{\left\Vert \mathsf{\hat{r}}_{%
\mathsf{j}}\right\Vert _{2}},\frac{\mathsf{\hat{r}}_{\mathsf{k}}}{\left\Vert 
\mathsf{\hat{r}}_{\mathsf{k}}\right\Vert _{2}}\right) _{2}\right\vert ^{2}<1.
\label{hilbertschmidtnorm}
\end{equation}%
Let us now consider the subspace of $l_{\mathbb{C}}^{2}$ defined by%
\begin{equation}
{\Large h}_{\mathsf{N}^{\ast }}:=\func{cl}\func{span}\left\{ \mathsf{\hat{r}}%
_{\mathsf{k}},\text{ }\mathsf{k}\in \left\{ \mathsf{N}^{\ast },\mathsf{N}%
^{\ast }\mathsf{+1},...\right\} \right\}  \label{hilbertsubspace}
\end{equation}%
where $\mathsf{N}^{\ast }$ is as in (\ref{hilbertschmidtnorm}). Let $M_{%
\mathsf{\hat{r},N}^{\ast }}$ stand for the infinite Gram matrix of the
normalized $\mathsf{\hat{r}}_{\mathsf{k}}$ with\textsf{\ }$\mathsf{k}$ as in
(\ref{hilbertsubspace}) and $\mathbb{I}_{\mathsf{N}^{\ast }}$ for the
identity operator on ${\Large h}_{\mathsf{N}^{\ast }}$. We then have%
\begin{equation*}
\left( \left( M_{\mathsf{\hat{r},N}^{\ast }}-\mathbb{I}_{\mathsf{N}^{\ast
}}\right) \mathsf{p}_{\mathsf{N}^{\ast }}\right) _{\mathsf{j}}\mathsf{=}%
\sum_{\mathsf{k=N}^{\ast },\mathsf{k\neq j}}^{+\infty }\left( \frac{\mathsf{%
\hat{r}}_{\mathsf{j}}}{\left\Vert \mathsf{\hat{r}}_{\mathsf{j}}\right\Vert
_{2}},\frac{\mathsf{\hat{r}}_{\mathsf{k}}}{\left\Vert \mathsf{\hat{r}}_{%
\mathsf{k}}\right\Vert _{2}}\right) _{2}p_{\mathsf{N}^{\ast },\mathsf{k}}
\end{equation*}%
where $\mathsf{j}\in \mathbb{N}^{+}$ with $\mathsf{j\geqslant N}^{\ast }$
and $\mathsf{p}_{\mathsf{N}^{\ast }}\in $ ${\Large h}_{\mathsf{N}^{\ast }}$.
In this manner it follows from (\ref{hilbertschmidtnorm}) that the
transformation $M_{\mathsf{\hat{r},N}^{\ast }}-\mathbb{I}_{\mathsf{N}^{\ast
}}$ is a Hilbert-Schmidt operator on ${\Large h}_{\mathsf{N}^{\ast }}$ with%
\begin{equation*}
\left\Vert M_{\mathsf{\hat{r},N}^{\ast }}-\mathbb{I}_{\mathsf{N}^{\ast
}}\right\Vert _{\mathsf{HS,N}^{\ast }}^{2}=\sum_{\mathsf{j,k\geqslant N}%
^{\ast }\mathsf{,j\neq k}}^{+\infty }\left\vert \left( \frac{\mathsf{\hat{r}}%
_{\mathsf{j}}}{\left\Vert \mathsf{\hat{r}}_{\mathsf{j}}\right\Vert _{2}},%
\frac{\mathsf{\hat{r}}_{\mathsf{k}}}{\left\Vert \mathsf{\hat{r}}_{\mathsf{k}%
}\right\Vert _{2}}\right) _{2}\right\vert ^{2}<1,
\end{equation*}%
where $\left\Vert .\right\Vert _{\mathsf{HS,N}^{\ast }}$ stands for the
Hilbert-Schmidt norm there. Therefore, we have \textit{a fortiori}%
\begin{equation*}
\left\Vert M_{\mathsf{\hat{r},N}^{\ast }}-\mathbb{I}_{\mathsf{N}^{\ast
}}\right\Vert _{\infty \mathsf{,N}^{\ast }}<1
\end{equation*}%
where $\left\Vert .\right\Vert _{\infty \mathsf{,N}^{\ast }\text{ }}$denotes
the usual sup-norm of the linear bounded operators on ${\Large h}_{\mathsf{N}%
^{\ast }}$. This proves that $M_{\mathsf{\hat{r},N}^{\ast }}=\mathbb{I}_{%
\mathsf{N}^{\ast }}-\left( \mathbb{I}_{\mathsf{N}^{\ast }}-M_{\mathsf{\hat{r}%
,N}^{\ast }}\right) $ is such an operator whose inverse is also bounded and
given by the corresponding Neumann series. Consequently, the $\mathsf{\hat{r}%
}_{\mathsf{k}}$ with $\mathsf{k}\in \left\{ \mathsf{N}^{\ast },\mathsf{N}%
^{\ast }\mathsf{+1},...\right\} $ constitute a basis of ${\Large h}_{\mathsf{%
N}^{\ast }}$ according to the fourth assertion of Theorem 2.1 in Chapter VI
of \cite{gohbergkrein}, that is, for every $\mathsf{p}_{\mathsf{N}^{\ast
}}\in $ ${\Large h}_{\mathsf{N}^{\ast }}$ we have the norm-convergent series
expansion%
\begin{equation*}
\mathsf{p}_{\mathsf{N}^{\ast }}=\dsum\limits_{\mathsf{k=N}^{\ast }}^{+\infty
}\left( \mathsf{p}_{\mathsf{N}^{\ast }}\mathsf{,\hat{s}}_{\mathsf{k}}\right)
_{2}\mathsf{\hat{r}}_{\mathsf{k}}.
\end{equation*}%
In order to get the result we want it remains to prove that we can complete
the basis just constructed with $\mathsf{\hat{r}}_{\mathsf{1}},...,\mathsf{%
\hat{r}}_{\mathsf{N}^{\ast }-1}$. Let $\mathsf{V}_{\mathsf{N}^{\ast }}$ be
the $\left( \mathsf{N}^{\ast }-1\right) $\textsf{-}dimensional subspace of $%
l_{\mathbb{C}}^{2}$ generated by these vectors. It follows from the
definitions of $\mathsf{V}_{\mathsf{N}^{\ast }}$, ${\Large h}_{\mathsf{N}%
^{\ast }}$ and from the completeness of all the $\mathsf{\hat{r}}_{\mathsf{k}%
}$ that%
\begin{equation*}
l_{\mathbb{C}}^{2}=\mathsf{V}_{\mathsf{N}^{\ast }}\oplus {\Large h}_{\mathsf{%
N}^{\ast }}
\end{equation*}%
as an algebraic direct sum. Therefore, any $\mathsf{p}\in l_{\mathbb{C}}^{2}$
may be written as%
\begin{equation*}
\mathsf{p=\dsum\limits_{\mathsf{k=1}}^{\mathsf{N}^{\ast }-1}}\gamma _{%
\mathsf{k}}\mathsf{\hat{r}}_{\mathsf{k}}\mathsf{+}\dsum\limits_{\mathsf{k=N}%
^{\ast }}^{+\infty }\left( \mathsf{p}_{\mathsf{N}^{\ast }}\mathsf{,\hat{s}}_{%
\mathsf{k}}\right) _{2}\mathsf{\hat{r}}_{\mathsf{k}}
\end{equation*}%
with some $\gamma _{\mathsf{k}}\in \mathbb{C}$ \ and $\mathsf{p}_{\mathsf{N}%
^{\ast }}\in $ ${\Large h}_{\mathsf{N}^{\ast }}$. Consequently, using the
biorthogonality properties%
\begin{equation*}
\left( \mathsf{\hat{r}}_{\mathsf{j}},\mathsf{\hat{s}}_{\mathsf{k}}\right)
_{2}=\delta _{\mathsf{j},\mathsf{k}}
\end{equation*}%
stemming from (\ref{renormalization}) and (\ref{biorthosequence}) we get%
\begin{equation*}
\left( \mathsf{p,\hat{s}}_{\mathsf{k}}\right) _{2}=\left\{ 
\begin{array}{c}
\gamma _{\mathsf{k}}\text{ \ for }\mathsf{k}\in \left\{ 1,...,\mathsf{N}%
^{\ast }-1\right\} , \\ 
\\ 
\left( \mathsf{p}_{\mathsf{N}^{\ast }}\mathsf{,\hat{s}}_{\mathsf{k}}\right)
_{2}\text{ \ \ for }\mathsf{k}\in \left\{ \mathsf{N}^{\ast },\mathsf{N}%
^{\ast }+1,...\right\}%
\end{array}%
\right.
\end{equation*}%
and thereby%
\begin{equation*}
\mathsf{p=}\dsum\limits_{\mathsf{k}=1}^{\mathsf{N}^{\ast }-1}\left( \mathsf{%
p,\hat{s}}_{\mathsf{k}}\right) _{2}\mathsf{\hat{r}}_{\mathsf{k}%
}+\dsum\limits_{\mathsf{k}=\mathsf{N}^{\ast }}^{\mathsf{+\infty }}\left( 
\mathsf{p}_{\mathsf{N}^{\ast }}\mathsf{,\hat{s}}_{\mathsf{k}}\right) _{2}%
\mathsf{\hat{r}}_{\mathsf{k}}=\dsum\limits_{\mathsf{k}=1}^{\mathsf{+\infty }%
}\left( \mathsf{p,\hat{s}}_{\mathsf{k}}\right) _{2}\mathsf{\hat{r}}_{\mathsf{%
k}},
\end{equation*}%
as desired. \ \ $\blacksquare $

\bigskip

\textsc{Remark.} According to the theorem we just referred to in the above
proof, the basis $\left( \mathsf{\hat{r}}_{\mathsf{k}}\right) \mathsf{\ }$of 
${\Large h}_{\mathsf{N}^{\ast }}$ with $\mathsf{k}\in \left\{ \mathsf{N}%
^{\ast },\mathsf{N}^{\ast }\mathsf{+1},...\right\} $ is actually a Riesz
basis, that is, one which may be obtained by a suitable deformation of an
orthonormal basis involving bounded invertible transformations. That notion
may be traced back to the original considerations set forth in Section
XXXVII in Chapter VII of \cite{paleywiener}, whose abstract version appears
in Section 86 in Chapter V of \cite{riesznagy}.

\bigskip

We are now ready for the following:

\bigskip

\textbf{Proof of the main theorem. }Owing to (\ref{renormalization}) and (%
\ref{biorthosequence}) we first have%
\begin{equation}
\dsum\limits_{\mathsf{k}=2}^{\mathsf{N}}\left( \mathsf{p,\hat{q}}_{\mathsf{k}%
}\right) _{2}\mathsf{\hat{p}}_{\mathsf{k}}=\dsum\limits_{\mathsf{k}=1}^{%
\mathsf{N}}\left( \mathsf{p,\hat{s}}_{\mathsf{k}}\right) _{2}\mathsf{\hat{r}}%
_{\mathsf{k}}+\left( \dsum\limits_{\mathsf{k}=2}^{\mathsf{N}}\left( \mathsf{%
p,\hat{q}}_{\mathsf{k}}\right) _{2}\right) \mathsf{\hat{p}}_{\mathsf{1}%
}-\left( \mathsf{p,\hat{s}}_{\mathsf{1}}\right) _{2}\mathsf{\hat{p}}_{%
\mathsf{1}}  \label{identitybis}
\end{equation}%
for each $\mathsf{p}\in l_{\mathbb{C}}^{2}$ and every $\mathsf{N}\in \mathbb{%
N}^{+}$, $\mathsf{N\geqslant 2}$. The issue being to establish (\ref{basis})
from (\ref{basisbis}), it is then necessary to prove the convergence of the
middle term on the right-hand side of (\ref{identitybis}) as $\mathsf{%
N\rightarrow +\infty }$ by an independent argument. For this it is
sufficient to take the inner product of the preceding equality by $\mathsf{%
\hat{q}}_{\mathsf{1}}$ given by (\ref{firstelement}). In so doing we obtain%
\begin{equation}
\dsum\limits_{\mathsf{k}=2}^{\mathsf{N}}\left( \mathsf{p,\hat{q}}_{\mathsf{k}%
}\right) _{2}=\left( \mathsf{p,\hat{s}}_{\mathsf{1}}\right) _{2}-\left(
\dsum\limits_{\mathsf{k}=1}^{\mathsf{N}}\left( \mathsf{p,\hat{s}}_{\mathsf{k}%
}\right) _{2}\mathsf{\hat{r}}_{\mathsf{k}},\mathsf{\hat{q}}_{\mathsf{1}%
}\right) _{2}  \label{identityter}
\end{equation}%
from the biorthogonality properties of Proposition 4, so that we have%
\begin{equation}
\dsum\limits_{\mathsf{k}=2}^{\mathsf{+\infty }}\left( \mathsf{p,\hat{q}}_{%
\mathsf{k}}\right) _{2}=\left( \mathsf{p,\hat{s}}_{\mathsf{1}}-\mathsf{\hat{q%
}}_{\mathsf{1}}\right) _{2}  \label{weakconvergence}
\end{equation}%
since (\ref{basisbis}) implies the convergence of the second term on the
right-hand side of (\ref{identityter}) to $(\mathsf{p,\hat{q}}_{\mathsf{1}%
})_{2}$. Therefore, letting $\mathsf{N\rightarrow +\infty }$ in (\ref%
{identitybis}) while using (\ref{weakconvergence}) we get%
\begin{equation*}
\dsum\limits_{\mathsf{k}=2}^{\mathsf{+\infty }}\left( \mathsf{p,\hat{q}}_{%
\mathsf{k}}\right) _{2}\mathsf{\hat{p}}_{\mathsf{k}}=\mathsf{p+\left( 
\mathsf{p,\hat{s}}_{\mathsf{1}}-\mathsf{\hat{q}}_{\mathsf{1}}\right) _{2}%
\hat{p}}_{\mathsf{1}}-\left( \mathsf{p,\hat{s}}_{\mathsf{1}}\right) _{2}%
\mathsf{\hat{p}}_{\mathsf{1}}
\end{equation*}%
in the sense of norm-convergence for every $\mathsf{p}\in l_{\mathbb{C}}^{2}$%
, which is (\ref{basis}). Finally, (\ref{semigrouprep}) is a direct
consequence of the continuity properties of the semigroup and of the
spectral properties of $A$. \ \ $\blacksquare $

\bigskip

The spectral decomposition of the main theorem now leads to the desired
description of the dynamics generated by (\ref{masterequationter}) and of
its ultimate behavior for large times, where we keep ordering the negative
eigenvalues of $A$ as $\nu _{\mathsf{k}}<\nu _{\mathsf{k+1}}$ for every $%
\mathsf{k}\in \left\{ 2,3,...\right\} $ and where $\left\Vert .\right\Vert
_{\infty \text{ \ }}$stands for the sup-norm of the linear bounded operators
on $l_{\mathbb{C}}^{2}$:

\bigskip

\textbf{Corollary. }\textit{Let us assume that the same hypotheses as in the
main theorem hold. Then the following statements are valid:}

\textit{(a) Let }$\mathsf{p}^{\ast }=\left( p_{\mathsf{m}}^{\ast }\right) _{%
\mathsf{m\in }\mathbb{N}^{+}}$ \textit{be any initial condition satisfying (%
\ref{probabilities}). Then we have}%
\begin{equation*}
\left( \exp \left[ \tau A\right] \mathsf{p}^{\ast }\right) _{\mathsf{m}%
}\geqslant 0,\text{ \ }\sum_{\mathsf{m=1}}^{\mathsf{+\infty }}\left( \exp %
\left[ \tau A\right] \mathsf{p}^{\ast }\right) _{\mathsf{m}}\text{\ }=1
\end{equation*}%
for every $\tau \in \left[ 0,+\infty \right) $.

\textit{(b) The Lyapunov exponent of the semigroup }$\exp \left[ \tau A%
\right] _{\tau \in \left[ 0,+\infty \right) }$ \textit{is given by}%
\begin{equation*}
\lim_{\tau \rightarrow +\infty }\frac{\ln \left\Vert \exp \left[ \tau A%
\right] \right\Vert _{\infty \text{ }}}{\tau }=0.
\end{equation*}

\textit{(c)} \textit{For each }$\mathsf{N}\in \mathbb{N}^{+},$ $\mathsf{%
N\geqslant 2}$, \textit{there exists a constant} $c_{\mathsf{N}}>0$\textit{\
such that for} \textit{every }$\mathsf{p}^{\ast }\in \left( \vee _{\mathsf{k}%
=\mathsf{N}+1}^{+\infty }E_{\nu _{\mathsf{k}}}(A^{\ast })\right) ^{\perp }$ 
\textit{we have the exponential decay estimate}%
\begin{equation*}
\left\Vert \exp \left[ \tau A\right] \mathsf{p}^{\ast }-\left( \mathsf{p}%
^{\ast }\mathsf{,\hat{q}}_{\mathsf{1}}\right) _{2}\mathsf{\hat{p}}%
_{1}\right\Vert _{2}\leqslant c_{\mathsf{N}}\exp \left[ -\tau \left\vert \nu
_{\mathsf{N}}\right\vert \right] \left\Vert \mathsf{p}^{\ast }\right\Vert
_{2}.
\end{equation*}%
\textit{In particular we have}%
\begin{equation}
\left\vert \left( \exp \left[ \tau A\right] \mathsf{p}^{\ast }\right) _{%
\mathsf{m}}-\left( \mathsf{p}^{\ast }\mathsf{,\hat{q}}_{\mathsf{1}}\right)
_{2}\mathsf{\hat{p}}_{_{1,\mathsf{m}}}\right\vert \leqslant c_{\mathsf{N}%
}\exp \left[ \tau \nu _{\mathsf{N}}\right] \left\Vert \mathsf{p}^{\ast
}\right\Vert _{2}  \label{expdecayestimate}
\end{equation}%
\textit{\ for each} $\mathsf{m}\in \mathbb{N}^{+}$ \textit{and every} $\tau
\in \left( 0,+\infty \right) $, \textit{where }$\mathsf{\hat{q}}_{\mathsf{1}%
} $\textit{\ is given by (\ref{firstelement}). }

\bigskip

\textbf{Proof. }The proof of Statement (a) follows immediately from (\ref%
{probabilities}), the continuity of $\tau \mapsto \exp \left[ \tau A\right] 
\mathsf{p}^{\ast }$ and the summation on both sides of (\ref%
{masterequationter}) over $\mathsf{m}\in \mathbb{N}^{+}$.

Statement (b) is a consequence of the very last part of Theorem 1 since%
\begin{equation*}
\lim_{\tau \rightarrow +\infty }\frac{\ln \left\Vert \exp \left[ \tau A%
\right] \right\Vert _{\infty }}{\tau }=\max_{\mathsf{k\in }\left\{
1,2,....\right\} }\nu _{\mathsf{k}}
\end{equation*}%
as an application of Theorem 4.1 in Chapter I of \cite{daleckiikrein}.

As for Statement (c), with $\mathsf{p}^{\ast }\in \left( \vee _{\mathsf{k}=%
\mathsf{N}+1}^{+\infty }E_{\nu _{\mathsf{k}}}(A^{\ast })\right) ^{\perp }$
we have%
\begin{equation}
\exp \left[ \tau A\right] \mathsf{p}^{\ast }-\left( \mathsf{p}^{\ast }%
\mathsf{,\hat{q}}_{\mathsf{1}}\right) _{2}\mathsf{\hat{p}}_{\mathsf{1}%
}=\dsum\limits_{\mathsf{k=2}}^{\mathsf{N}}\left( \mathsf{p}^{\ast }\mathsf{,%
\hat{q}}_{\mathsf{k}}\right) _{2}\exp \left[ \tau \nu _{\mathsf{k}}\right] 
\mathsf{\hat{p}}_{\mathsf{k}}  \label{truncatedexpansion}
\end{equation}%
from (\ref{semigrouprep}) since then $\left( \mathsf{p}^{\ast }\mathsf{,\hat{%
q}}_{\mathsf{k}}\right) _{2}=0$ for every $\mathsf{k\geqslant N+1}$, so that
the estimate%
\begin{equation*}
\left\Vert \exp \left[ \tau A\right] \mathsf{p}^{\ast }-\left( \mathsf{p}%
^{\ast }\mathsf{,\hat{q}}_{\mathsf{1}}\right) _{2}\mathsf{\hat{p}}_{\mathsf{1%
}}\right\Vert _{2}\leqslant c_{\mathsf{N}}\exp \left[ -\tau \left\vert \nu _{%
\mathsf{N}}\right\vert \right] \left\Vert \mathsf{p}^{\ast }\right\Vert _{2}
\end{equation*}%
indeed holds with an obvious choice for $c_{\mathsf{N}}$, which immediately
leads to (\ref{expdecayestimate}). \ \ $\blacksquare $

\bigskip

\textsc{Remarks.} (1) Since $\lim_{\mathsf{k}\rightarrow +\infty }\nu _{%
\mathsf{k}}=0$ as a consequence of the compactness of $A$ (see, e.g.,
Theorem 7.1 in Chapter VII of \cite{conway}), Statement (c) of the corollary
is very different from the corresponding assertions which one might get in
finite-dimensional situations, as there is no spectral gap between $\nu
_{1}=0$ and the remaining eigenvalues. Thus, our conclusion is that there is
a large supply of initial conditions to choose from for any $\mathsf{N}\in 
\mathbb{N}^{+},$ $\mathsf{N\geqslant 2}$, such that the corresponding
solutions to (\ref{masterequationter}) stabilize exponentially rapidly to $%
\gamma \mathsf{p}_{\beta ,\mathsf{Gibbs}}$ for some $\gamma \in \mathbb{C}$.
Indeed, it is useful to recall here that $\mathsf{\hat{p}}_{1}$ as given by (%
\ref{generator}) with $\mathsf{k}=1$ is a scalar multiple of $\mathsf{p}%
_{\beta ,\mathsf{Gibbs}}$. Furthermore the larger $\mathsf{N}$ is, the
larger that supply becomes but this is at the expense of having $\lim_{%
\mathsf{N}\rightarrow +\infty }\exp \left[ -\tau \left\vert \nu _{\mathsf{N}%
}\right\vert \right] =1$. Therefore, there is a clear trade-off between the
dimension of $\left( \vee _{\mathsf{k}=\mathsf{N}+1}^{+\infty }E_{\nu _{%
\mathsf{k}}}(A^{\ast })\right) ^{\perp }$ and the rate of decay.

(2) The choice of the initial condition $\mathsf{p}^{\ast }$ in Statement
(c) of Corollary 1 was motivated by the desire to generate the truncated
expansion (\ref{truncatedexpansion}) in order to obtain the exponential rate
of decay in (\ref{expdecayestimate}). A natural question is therefore
whether that truncation technique is really necessary to get at least some
type of convergence toward a multiple of $\mathsf{p}_{\beta ,\mathsf{Gibbs}}$%
, in spite of the fact that $\nu _{1}=0$ is an accumulation point of $\sigma
(A)$. So far we have been able to show that%
\begin{equation}
\lim_{\tau \rightarrow +\infty }\left\Vert \exp \left[ \tau A\right] \mathsf{%
p}^{\ast }-\left( \mathsf{p}^{\ast }\mathsf{,\hat{q}}_{\mathsf{1}}\right)
_{2}\mathsf{\hat{p}}_{\mathsf{1}}\right\Vert _{2}=0  \label{convergenceter}
\end{equation}%
for every $\mathsf{p}^{\ast }\in l_{\mathbb{C}}^{1}\subset l_{\mathbb{C}%
}^{2} $, a very partial answer indeed as (\ref{convergenceter}) does not
provide rates of decay, nor does it say whether it holds for all $\mathsf{p}%
^{\ast }\in l_{\mathbb{C}}^{2}$ and not merely for all $\mathsf{p}^{\ast
}\in l_{\mathbb{C}}^{1}$. Its proof is based on (\ref{semigrouprep}) written
as 
\begin{equation*}
\exp \left[ \tau A\right] \mathsf{p}^{\ast }-\left( \mathsf{p}^{\ast }%
\mathsf{,\hat{q}}_{\mathsf{1}}\right) _{2}\mathsf{\hat{p}}_{\mathsf{1}%
}=\dsum\limits_{\mathsf{k=2}}^{\mathsf{+\infty }}\left( \mathsf{p}^{\ast }%
\mathsf{,\hat{q}}_{\mathsf{k}}\right) _{2}\exp \left[ \tau \nu _{\mathsf{k}}%
\right] \mathsf{\hat{p}}_{\mathsf{k}},
\end{equation*}%
and on a very careful estimate of the $l_{\mathbb{C}}^{2}$-norm of the
right-hand side which rests on some of the inequalities proved in the core
of this section. We omit the details.

(3) In many applications the real sequence $\left( \lambda _{\mathsf{m}%
}\right) _{\mathsf{m}\in \mathbb{N}^{+}}$ used in the above considerations
represents the pure point spectrum of some differential operator, typically
a Hamiltonian operator in Quantum Mechanics. In this setting the
coefficients $r_{\mathsf{m,n}}$ given by (\ref{transitionrate}) represent
the transition rates between the corresponding eigenstates labeled $\mathsf{n%
}$ and $\mathsf{m}$, respectively, and equations of the form (\ref%
{masterequation}) may be used to describe the transient regime of a system
that approaches thermodynamical equilibrium by using entropy production
arguments, as was done at a formal level in Part D of Section II of \cite%
{tomeoliveira} to which we refer the reader for details. Furthermore there
are plenty of operators whose pure point spectrum satisfies (\ref%
{partitionfunction}), together with the conditions of all the propositions
and theorems of this section, for instance the Hamiltonian operator
describing the quantum harmonic oscillator. For an extension of the use of
master equations as they relate to the investigation of physical or chemical
systems in contact with reservoirs, we refer the reader again to \cite{haake}
and \cite{mozgunovlidar}, and also to Sections 10.4 and 10.5 in Chapter 10
of \cite{davies} where a functional-analytical treatment is carried out, and
where many references to the physics litterature are given.

(4) Entropy production arguments may also be used to generalize the
investigations carried out in \cite{vuillermot} regarding a special class of
stochastic processes, to which we can associate time-dependent entropy
functionals of the form 
\begin{equation*}
\mathsf{S}(\tau )=\sum_{\mathsf{m}=1}^{+\infty }p_{\mathsf{m}}(\tau )\ln 
\frac{1}{p_{\mathsf{m}}(\tau )}
\end{equation*}%
where $p_{\mathsf{m}}(\tau )$ is a solution to (\ref{masterequation}). The
presentation of the related results is deferred to a separate publication.

\bigskip

We complete this article with the two appendices we alluded to in the
introduction.

\bigskip

\textbf{Appendix A. On the holomorphic continuation of the function given by
(\ref{function}).}

The following result holds:

\bigskip

\textbf{Proposition A.1. }\textit{Let} $\hat{f}:\mathbb{C\setminus }\left\{
0,\text{ }-b_{\mathsf{m}},\text{ }\mathsf{m}\in \mathbb{N}^{+}\right\}
\mapsto $ $\mathbb{C}$ \textit{be the function defined by}%
\begin{equation}
\hat{f}\left( \nu \right) :=\dsum\limits_{\mathsf{m=1}}^{\mathsf{+\infty }}%
\frac{\exp \left[ -\alpha \beta \lambda _{\mathsf{m}}\right] }{\nu +b_{%
\mathsf{m}}}  \label{functionbis}
\end{equation}%
\textit{where }$b_{\mathsf{m}}$\textit{\ is given by (\ref{shorthand})}. 
\textit{Then }$\hat{f}$ \textit{is holomorphic throughout its domain and we
have}%
\begin{equation}
\hat{f}^{\prime }\left( \nu \right) =-\dsum\limits_{\mathsf{m=1}}^{\mathsf{%
+\infty }}\frac{\exp \left[ -\alpha \beta \lambda _{\mathsf{m}}\right] }{%
\left( \nu +b_{\mathsf{m}}\right) ^{2}}.  \label{derivation}
\end{equation}

\bigskip

\textbf{Proof. }The absolute convergence of (\ref{functionbis}) in every
point of its domain is proved as for the function given by (\ref{function}).
Now let $\nu _{\ast }\in \mathbb{C\setminus }\left\{ 0,\text{ }-b_{\mathsf{m}%
},\text{ }\mathsf{m}\in \mathbb{N}^{+}\right\} $ be arbitrary, and let%
\begin{equation*}
\hat{f}_{\mathsf{N}}\left( \nu \right) :=\dsum\limits_{\mathsf{m=1}}^{%
\mathsf{N}}\frac{\exp \left[ -\alpha \beta \lambda _{\mathsf{m}}\right] }{%
\nu +b_{\mathsf{m}}}
\end{equation*}%
be the holomorphic partial sums of (\ref{functionbis}). In order to prove
that $\hat{f}$ is holomorphic at $\nu _{\ast }$ it is sufficient to prove
that $\hat{f}_{\mathsf{N}}\rightarrow \hat{f}$ uniformly on a compact disk
of sufficiently small radius $R_{\nu _{\ast }}$ centered at $\nu _{\ast }$.
To this end we consider%
\begin{equation*}
\mathbb{D}_{R_{\nu _{\ast }}}(\nu _{\ast })=\left\{ \nu \in \mathbb{C}%
:\left\vert \nu -\nu _{\ast }\right\vert \leq R_{\nu _{\ast }}\right\} 
\end{equation*}%
with $0<R_{\nu _{\ast }}<\left\vert \nu _{\ast }\right\vert $ and show that 
\begin{equation}
\sup_{\nu \in \mathbb{D}_{R_{\nu _{\ast }}}(\nu _{\ast })}\left\vert \hat{f}%
\left( \nu \right) -\hat{f}_{\mathsf{N}}\left( \nu \right) \right\vert \leq 
\frac{2}{\left\vert \nu _{\ast }\right\vert -R_{\nu _{\ast }}}\dsum\limits_{%
\mathsf{m=N+1}}^{\mathsf{+\infty }}\exp \left[ -\alpha \beta \lambda _{%
\mathsf{m}}\right]   \label{uniformconvergence}
\end{equation}%
for $\mathsf{N}$ sufficiently large, which indeed implies the desired
convergence by virtue of (\ref{partitionfunction}). Owing to the choice of $%
R_{\nu _{\ast }}$ we first have%
\begin{eqnarray}
&&\left\vert \hat{f}\left( \nu \right) -\hat{f}_{\mathsf{N}}\left( \nu
\right) \right\vert   \notag \\
&\leq &\frac{1}{\left\vert \nu _{\ast }\right\vert -\left\vert \nu -\nu
_{\ast }\right\vert }\dsum\limits_{\mathsf{m=N+1}}^{\mathsf{+\infty }}\frac{%
\exp \left[ -\alpha \beta \lambda _{\mathsf{m}}\right] }{\left\vert 1+\frac{%
b_{\mathsf{m}}}{\nu }\right\vert }  \label{estimate} \\
&\leq &\frac{1}{\left\vert \nu _{\ast }\right\vert -R_{\nu _{\ast }}}%
\dsum\limits_{\mathsf{m=N+1}}^{\mathsf{+\infty }}\frac{\exp \left[ -\alpha
\beta \lambda _{\mathsf{m}}\right] }{\left\vert 1+\frac{b_{\mathsf{m}}}{\nu }%
\right\vert }.  \notag
\end{eqnarray}%
Furthermore, since $b_{\mathsf{m}}\rightarrow 0$ as $\mathsf{m\rightarrow
+\infty }$ there exists $\mathsf{N}_{\nu _{\ast }}\in \mathbb{N}^{+}$ such
that 
\begin{equation*}
b_{\mathsf{m}}\leq \frac{\left\vert \nu _{\ast }\right\vert -R_{\nu _{\ast }}%
}{2}
\end{equation*}%
for every $\mathsf{m\geq N}_{\nu _{\ast }}$, which implies the estimate%
\begin{equation}
\left\vert 1+\frac{b_{\mathsf{m}}}{\nu }\right\vert \geq 1-\frac{b_{\mathsf{m%
}}}{\left\vert \nu _{\ast }\right\vert -R_{\nu _{\ast }}}\geq \frac{1}{2}
\label{estimatebis}
\end{equation}%
uniformly in $\nu $. Using (\ref{estimatebis}) in (\ref{estimate}) with $%
\mathsf{N}\geq $ $\mathsf{N}_{\nu _{\ast }}$ then leads to (\ref%
{uniformconvergence}), so that $\hat{f}$ is holomorphic at $\nu _{\ast }$
and hence in $\mathbb{C\setminus }\left\{ 0,\text{ }-b_{\mathsf{m}},\text{ }%
\mathsf{m}\in \mathbb{N}^{+}\right\} $ since $\nu _{\ast }$ was arbitrary.

Similar estimates allow one to prove that $\hat{f}_{\mathsf{N}}^{\prime
}\rightarrow \hat{f}^{\prime }$ uniformly on disks of sufficiently small
radii where%
\begin{equation*}
\hat{f}_{\mathsf{N}}^{\prime }\left( \nu \right) =-\dsum\limits_{\mathsf{m=1}%
}^{\mathsf{N}}\frac{\exp \left[ -\alpha \beta \lambda _{\mathsf{m}}\right] }{%
\left( \nu +b_{\mathsf{m}}\right) ^{2}},
\end{equation*}%
eventually establishing (\ref{derivation}) (see, e.g., Section 3.5 in
Chapter 3 of \cite{greenkrantz}). \ \ $\blacksquare $

\bigskip

Since $f$ given by (\ref{function}) is the restriction of $\hat{f}$ \ to $%
\left( -\infty ,0\right) \setminus \left\{ -b_{\mathsf{m}},\text{ }\mathsf{m}%
\in \mathbb{N}^{+}\right\} $, the preceding result justifies \textit{a
posteriori }the few\textit{\ }properties of that function we used in the
proof of Statement (c) of Theorem 1 regarding the localization of the
eigenvalues $\nu _{\mathsf{k}}$.

\bigskip

\textbf{Appendix B. A simple characterization of the eigenvectors of }$%
A^{\ast }$

In this short appendix we wish to characterize in a geometric way the
sequence $\left( \mathsf{\hat{q}}_{\mathsf{k}}\right) _{\mathsf{k}\in 
\mathbb{N}^{+}}$ biorthogonal to $\left( \mathsf{\hat{p}}_{\mathsf{k}%
}\right) _{\mathsf{k}\in \mathbb{N}^{+}}$ constructed in the proof of
Proposition 4. Let us consider the direct sum decomposition%
\begin{equation*}
\ l_{\mathbb{C}}^{2}=\dbigvee\limits_{\mathsf{k=1,k\neq j}}^{+\infty }E_{\nu
_{\mathsf{k}}}(A)\oplus \left( \dbigvee\limits_{\mathsf{k=1,k\neq j}%
}^{+\infty }E_{\nu _{\mathsf{k}}}(A)\right) ^{\perp }
\end{equation*}%
for every $\mathsf{j}\in \mathbb{N}^{+}$, where the first space on the
right-hand side stands for the closed linear hull of $\cup _{\mathsf{%
k=1,k\neq j}}^{+\infty }E_{\nu _{\mathsf{k}}}(A)$. We then consider the
orthogonal projection%
\begin{equation*}
Q_{\mathsf{j}}:l_{\mathbb{C}}^{2}\mapsto \left( \dbigvee\limits_{\mathsf{%
k=1,k\neq j}}^{+\infty }E_{\nu _{\mathsf{k}}}(A)\right) ^{\perp },
\end{equation*}%
and remark that $Q_{\mathsf{j}}\mathsf{\hat{p}}_{\mathsf{j}}\neq 0$ for
every $\mathsf{j}$. Indeed, $Q_{\mathsf{j}^{\ast }}\mathsf{\hat{p}}_{\mathsf{%
j}^{\ast }}=0$ for at least one\textsf{\ }$\mathsf{j}^{\ast }$ would mean
that%
\begin{equation*}
\mathsf{\hat{p}}_{\mathsf{j}^{\ast }}\in \ker Q_{\mathsf{j}^{\ast
}}=\dbigvee\limits_{\mathsf{k=1,k\neq j}^{\ast }}^{+\infty }E_{\nu _{\mathsf{%
k}}}(A),
\end{equation*}%
which in turn would entail the relation%
\begin{equation}
\func{span}\cup _{\mathsf{k=1}}^{+\infty }E_{\nu _{\mathsf{k}}}(A)\subseteq
\dbigvee\limits_{\mathsf{k=1,k\neq j}^{\ast }}^{+\infty }E_{\nu _{\mathsf{k}%
}}(A).  \label{inclusionquarto}
\end{equation}%
But by taking the closure of (\ref{inclusionquarto}) in $l_{\mathbb{C}}^{2}$
and by using (\ref{completeness}) we would then obtain%
\begin{equation*}
l_{\mathbb{C}}^{2}\subseteq \dbigvee\limits_{\mathsf{k=1,k\neq j}^{\ast
}}^{+\infty }E_{\nu _{\mathsf{k}}}(A)\subset l_{\mathbb{C}}^{2},
\end{equation*}%
a contradiction. We may therefore consider the sequence $\left( \mathsf{\hat{%
t}}_{\mathsf{j}}\right) _{\mathsf{j}\in \mathbb{N}^{+}}$ given by%
\begin{equation}
\mathsf{\hat{t}}_{\mathsf{j}}:=\left\Vert Q_{\mathsf{j}}\mathsf{\hat{p}}_{%
\mathsf{j}}\right\Vert _{2}^{-2}Q_{\mathsf{j}}\mathsf{\hat{p}}_{\mathsf{j}},
\label{newsequence}
\end{equation}%
from which we easily see that 
\begin{equation*}
\left( \mathsf{\hat{p}}_{\mathsf{j}},\mathsf{\hat{t}}_{\mathsf{k}}\right)
_{2}=\delta _{\mathsf{j,k}}
\end{equation*}%
for all $\mathsf{j,k}\in \mathbb{N}^{+}$. Consequently, the sequence $\left( 
\mathsf{\hat{t}}_{\mathsf{j}}\right) _{\mathsf{j}\in \mathbb{N}^{+}}$ is
biorthogonal to $\left( \mathsf{\hat{p}}_{\mathsf{k}}\right) _{\mathsf{k}\in 
\mathbb{N}^{+}}$ and the preceding relation together with (\ref%
{biorthogonalitybis}) immediately imply that 
\begin{equation}
\mathsf{\hat{t}}_{\mathsf{j}}=\mathsf{\hat{q}}_{\mathsf{j}}  \label{equality}
\end{equation}%
for every $\mathsf{j}\in \mathbb{N}^{+}$ since the $\mathsf{\hat{p}}_{%
\mathsf{j}}$ form a complete system in $l_{\mathbb{C}}^{2}$. Thus the
sequence $\left( \mathsf{\hat{q}}_{\mathsf{k}}\right) _{\mathsf{k}\in 
\mathbb{N}^{+}}$ of Proposition 4 is the unique sequence biorthogonal to $%
\left( \mathsf{\hat{p}}_{\mathsf{k}}\right) _{\mathsf{k}\in \mathbb{N}^{+}}$%
, and (\ref{newsequence}) with (\ref{equality}) provide a simple geometric
characterization of the eigenvectors of $A^{\ast }$.

\bigskip

\textbf{Acknowledgements. }The second author would like to thank the Funda%
\c{c}\~{a}o para a Ci\^{e}ncia e a Tecnologia (FCT) of the Portuguese
Government for its financial support under grant PDTC/MAT-STA/0975/2014.


\begin{thebibliography}{99}
\bibitem{conway} \textsc{Conway, J. B.,} \textit{A Course in Functional
Analysis,} Graduate Texts in Mathematics \textbf{96}, Springer Verlag, New
York (1990).

\bibitem{daleckiikrein} \textsc{Dalecki\u{\i}, Ju. L., Krein, M. G.,}\textit{%
\ Stability of Solutions of Differential Equations in Banach Space,}
Translations of Mathematical Monographs \textbf{43}, American Mathematical
Society, Providence (1974).

\bibitem{davies} \textsc{Davies, E. B.,} \textit{Quantum Theory of Open
Systems, }Academic Press, New York (1976).

\bibitem{gelfandvilenkin} \textsc{Gelfand, I. M., Vilenkin, N.Ya.,} \textit{%
Generalized Functions,} Vol. 4: \textit{Applications of Harmonic Analysis,}
Academic Press, New York (1964).

\bibitem{gohbergkrein} \textsc{Gohberg, I. C., Krein, M. G., }\textit{%
Introduction to the Theory of Linear Nonselfadjoint Operators in Hilbert
Space,} Translations of Mathematical Monographs \textbf{18}, American
Mathematical Society, Providence (1969).

\bibitem{greenkrantz} \textsc{Greene, R. E., Krantz, S. G., }\textit{%
Function Theory of One Complex Variable,} Graduate Studies in Mathematics 
\textbf{40}, American Mathematical Society, Providence (2006).

\bibitem{haake} \textsc{Haake, F.,} \textit{Statistical treatment of open
systems by generalized master equations,} in: Springer Tracts in Modern
Physics \textbf{66}, Springer, New York (1973).

\bibitem{keldys} \textsc{Keldy\u{s}, M. V.,} \textit{On the characteristic
values and characteristic functions of certain classes of non self-adjoint
equations,} Doklady Akad. Nauk SSSR \textbf{77,} 11-14 (1951).

\bibitem{mozgunovlidar} \textsc{Mozgunov, E., Lidar, D., }\textit{Completely
positive master equation for arbitrary driving and small level spacing,}
Quantum \textbf{4}, 227-289 (2020).

\bibitem{paleywiener} \textsc{Paley, R. E. A. C., Wiener, N.,} \textit{%
Fourier Transforms in the Complex Plane,} American Mathematical Society
Colloquium Publications XIX, American Mathematical Society, New York, (1934).

\bibitem{riesznagy} \textsc{Riesz, F., Nagy, B. SZ., }\textit{Functional
Analysis,} Dover Books in Mathematics, Dover, (1990).

\bibitem{schnakenberg} \textsc{Schnakenberg, J.,} \textit{Network theory of
microscopic and macroscopic behavior of master equation systems,} Reviews of
Modern Physics \textbf{48}, 571-585 (1976).

\bibitem{tomeoliveira} \textsc{Tom\'{e}, T., de Oliveira, M. J., }\textit{%
Stochastic approach to equilibrium and nonequilibrium thermodynamics,}
Physical Review E \textbf{91}, 042140 (2015).

\bibitem{tomeoliveirabis} \textsc{Tom\'{e}, T., de Oliveira, M. J.,} \textit{%
Stochastic thermodynamics and entropy production of chemical reaction
systems, }Journal of Chemical Physics \textbf{148}, 224104 (2018).

\bibitem{vankampen} \textsc{van Kampen, N. G.,} \textit{Stochastic Processes
in Physics and Chemistry,} Elsevier Science Publishers B. V., Amsterdam
(1981).

\bibitem{vinberg} \textsc{Vinberg, E. B.,} \textit{A Course in Algebra,}
Graduate Studies in Mathematics \textbf{56}, American Mathematical Society,
Providence, Rhode Island (2003).

\bibitem{vuillermot} \textsc{Vuillermot, P.-A., }\textit{On Bernstein
processes of maximal entropy,} Stochastic Analysis and Applications \textbf{%
38}, 886-908 (2020).
\end{thebibliography}
\end{document}